\documentclass[review]{elsarticle}

\usepackage[cmex10]{amsmath}
\usepackage{amsfonts}
\usepackage{algorithm}
\usepackage[noend]{algpseudocode}
\usepackage{graphicx}
\usepackage{caption}
\usepackage{subcaption}
\usepackage{pdflscape}
\usepackage{color}
\usepackage{xr}
\newcommand*\rfrac[2]{{}^{#1}\!/_{#2}}

\usepackage[margin=2.5cm]{geometry}

\DeclareMathOperator*{\argmax}{arg\,max}

\journal{arXiv}

\usepackage{numcompress}

\begin{document}

\begin{frontmatter}

\title{An Evaluation of Sparse Inverse Covariance Models for Group Functional Connectivity in Molecular Imaging}

%\author[IH]{D.B..~Keator\corref{cor1}\fnref{fn1}}
\author[IH]{D.B..~Keator\corref{cor1}\fnref{fn1}}
\author[DBH]{A.~Ihler\fnref{fn2}}

\cortext[cor1]{Corresponding author}
\fntext[fn1] {Email Address: dbkeator@uci.edu (D.B. Keator)}
\fntext[fn2] {Email Address: aihler@uci.edu (A. Ihler)}

\address[IH]{University of California, Irvine, Irvine Hall rm. 163 - ZOT 3960, Irvine CA. 92697-3435}
\address[DBH]{University of California, Irvine, Donald Bren Hall, Irvine CA. 92697-3435}

\begin{abstract}
Evaluating the functional relationships between brain regions measured with neuroimaging provides insight into how the brain is sharing information at a macro scale.  Many functional connectivity methods have been developed for dynamic imaging modalities such as functional MRI (fMRI), while less work has focused on models for static molecular imaging techniques such as FDG-PET and Tc-99m HMPAO SPECT across groups of individuals.  In this work we provide a quantitative assessment of how well three functional connectivity models based on sparse inverse covariance estimation can accurately recover gold standard connectivity patterns across multiple cohorts and data set sizes.  We compare the accuracies of learning regularized inverse covariance matrices across cohorts independently with those learned using two different group-based regularization models.  By using large cohorts of SPECT scans, we are able to provide a quantitative assessment of the accuracy of the models in recovering the gold standard functional connectivity patterns and connectivity strengths across a range of typical data set sizes in neuroimaging studies.          
\end{abstract}
\begin{keyword}
\texttt{Functional connectivity}\sep SPECT \sep PET \sep graphical model
%\MSC[2010] 00-01\sep  99-00
\end{keyword}

\end{frontmatter}

%\linenumbers

\section{Introduction}
Neuroimaging modalities are routinely used to evaluate regional differences in brain function and connectivity.   Neuroimaging provides researchers with a method to evaluate in-vivo brain function, where various statistics relating activity between spatially distributed regions are used as indicators of functional connectedness \cite{Friston:1994ct}.  Many studies have shown changes in functional connectivity, as assessed with neuroimaging, across genders \cite{Ingalhalikar:2014io,Biswal:2010fj} and in psychiatric illnesses such as schizophrenia \cite{McGuire:1996jz,Micheloyannis:2006hv} and Alzheimer's Disease \cite{Anonymous:IlbMPcJj,Wang:2007hh}  among others.  The majority of studies and methods of evaluating functional connectivity have focused on dynamic imaging modalities such as functional magnetic resonance imaging (fMRI) and 15O-H$_{2}$0 Positron Emission Tomography (PET).  Fewer studies have evaluated functional connectivity in static molecular imaging modalities such as $^{18}$F-FDG PET and Tc-99m HMPAO Single Photon Emission Computed Tomography (SPECT) across groups of subjects.   \citet{Anonymous:IlbMPcJj}  compares functional connectivity across cohorts using static PET; however, their results give little information on how well the models recover the true connectivity profiles of the populations tested.  Careful validation of connectivity models is necessary, prior to their use and interpretation in real data \cite{smith2011network}. 
\\
\\
A relatively simple method of interrogating functional connectivity in neuroimaging is through inter-regional and/or voxel-wise correlations \cite{Rogers:2007gx,Clark:1984uq}.   This method typically consists of cross-correlating the mean functional activity for each region of interest (ROI) and applying a threshold to remove values close to zero. Unfortunately, it has been shown that the sample correlation matrix is often a poor estimator of the population correlation matrix and can result in incorrect conclusions based on estimation error \cite{varoquaux2013learning} \cite{smith2011network}.  Further, correlations between two ROIs does not imply a direct functional connection and thus impedes our ability to interpret our results with respect to known brain circuits.   A more interpretable statistic, that has been shown to estimate the true network \cite{smith2011network}, is the partial correlation: the statistical relationship between pairs of brain regions after conditioning on the others.  If the brain regions are jointly Gaussian, then we can find regions with strong partial correlation by examining the non-zero entries in the scaled inverse covariance matrix of the Gaussian distribution, fit to the observed data. Unfortunately, in typical neuroimaging studies, we do not have enough data to accurately compute the inverse of the covariance matrix among brain regions; therefore, some form of regularization is needed.  To estimate the inverse covariance matrix and thus the partial correlations consistent with our observed data, we can use regularized maximum likelihood methods  \cite{Tibshirani:1996bc,Banerjee:2008ud}.  Although we have methods available, we lack data on how well these classes of models can recover the true connectivity in group-based molecular imaging and how sample sizes effect accuracy of the results.  
\\
\\
In this study we compare three models for learning functional connectivity: one where the connectivity profiles are learned independently for each cohort and two that share information between cohorts.  In settings with low sample sizes, we hypothesize that models which share connectivity information across cohorts will perform better in recovering correct connections.   Our contribution is a quantitative assessment of how well the functional connectivity models can accurately recover a gold standard connectivity pattern across a range of typical data set sizes in neuroimaging studies based on a very large sample of SPECT scans.  To our knowledge, this is the first study evaluating the accuracy of recovered connectivity profiles by sample size in group-based static molecular imaging.  The rest of the manuscript is organized as follows.  In the section \ref{methods}, we describe three models for learning sparse inverse covariance matrices which have been applied to similar problems.  Next, we describe the gold standard connectivity data used to quantitatively evaluate the inverse covariance models.   We conclude section \ref{methods} with a summary of the model.   In section \ref{results}, we present a detailed comparison of the sparse inverse covariance models with respect to a gold standard connectivity profile, as a function of sample size using two different methods of setting the model regularization coefficients.  We conclude the manuscript with a discussion of the results and plans for future work.

\section{Methods}
\label{methods}
In this section we describe three sparse inverse covariance models for learning group-based functional connectivity patterns between pairs of regions, one of which learns the patterns independently for each cohort and two models which share connectivity information across cohorts.  We anticipate the models that share connectivity information across cohorts will improve our prediction accuracy in settings with small sample sizes.  We conclude the methods section with a description of the gold standard dataset we use to quantitatively evaluate the performance of each connectivity model.   
\\
\\
\subsection{Sparse Inverse Covariance Estimation}
\label{SICE}
Ultimately we are interested in learning about how the brain is functionally connected at the group level across different experimental conditions and/or cohorts.  In learning about the functional connectivity in our data, we are presented with a few challenges.  First, the brain is composed of approximately 100 billion neurons, with each neuron being connected with up to 10,000 other neurons, potentially forming 1 trillion synaptic connections.  Even at the scale of  neuroimaging, we are likely to find some association between any pair of brain regions.  We need a principled approach that focuses our attention on the most relevant associations in a way that can be related to known brain circuits.  Next, we are constrained by small sample sizes.  Often, particularly in rare diseases, it is difficult to find large groups of subjects, making it more difficult to learn about whole-brain functional connectivity.  Fortunately, there are models available that can help us learn about functional connectivity in light of these problems, yet there is little data on how accurately the models perform when applied to molecular brain imaging data.  
\\
\\
Sparse inverse covariance estimation models provide a structured approach of estimating pairwise connections between variables in settings with small numbers of samples.   Entries in the inverse covariance matrix (also called the precision matrix) of a Gaussian model, correspond to the pairwise statistical relationships between variables conditioned on all the other variables in the model.  Values of zero in the inverse covariance matrix (i.e., zero partial correlation) imply conditional independence of the associated variables \cite{marrelec2006partial}.  Alternatively, strong partial correlations are suggestive of pairwise interactions, helping us to interpret results with respect to known brain circuits. In contrast, bivariate correlations provide pairwise associations but do not account for all the other brain regions and their potential effect on the correlation; although, using both partial and bivariate correlations together when interpreting results is well advised.  Learning inverse covariance matrices in settings with small data set sizes and a large number of variables is difficult and some form of regularization is needed to both improve prediction accuracy and aid in interpretation, focusing on a smaller subset that exhibit the strongest effects \cite{Tibshirani:1996bc}.   In section \ref{glasso} we describe Graphical Lasso (GL), a popular technique for learning sparse inverse covariance matrices.  GL uses regularization to improve prediction accuracy and interpretability in settings with small sample sizes relative to the number of variables \cite{Tibshirani:1996bc}; however, in many experiments the groups being compared are similar, and we would expect some of the functional connectivity relationships are shared across cohorts.  We would like to use this observation to help improve our prediction accuracy.  In section \ref{joint_glasso} we describe the Fused Graphical Lasso (FGL) and Group Graphical Lasso (GGL) models proposed by \citet{Danaher:2013fe} based on work from \citet{tibshirani2005sparsity} for jointly estimating sparse precision matrices across cohorts.  Our contribution is a quantitative analysis of how well these models perform in recovering a gold standard connectivity profile as a function of sample size in static molecular imaging.  

\subsubsection{Graphical Lasso}
\label{glasso}
The least absolute shrinkage and selection operator (lasso) proposed by \citet{Tibshirani:1996bc} is a method for penalized regression that shrinks some parameter estimates and sets others to zero.  Shrinking parameter estimates controls over fitting and results in models that have better predictive accuracy.   Setting some parameters to zero provides variable selection,  focusing on the more relevant features of the data, resulting in more interpretable modeling results.  \citet{Banerjee:2008ud} applied the lasso penalty to learning sparse undirected graphical models in a multivariate Gaussian setting and developed a block-wise interior point algorithm which they noted is equivalent to iteratively solving lasso problems.   \citet{Friedman:2008df} pursued this observation and developed the Graphical Lasso (GL) algorithm.  The learning problem is to maximize the penalized log-likelihood over all positive definite matrices $\Phi$ given by:
 \begin{equation}
 \label{mle_glasso}
\hat{\Sigma}^{-1} =  \argmax_{\Phi \succ 0} \ \log \det \Phi - \operatorname{tr.}(S\Phi) - \lambda \|\Phi\|_1
 \end{equation}
 where $\hat{\Sigma}^{-1}$ is the estimate of the sparse precision matrix, $S$ is the empirical covariance matrix of the data computed using regions of interest (see  \ref{clustering}), where entry $S_{i,j}$ is the covariance between regions $i$ and $j$ across all subjects in the group, and $\|\Phi\|_1$ is the $L_1$ norm, the sum of the absolute values of the elements of the current estimate of $\Phi$.  The non-negative coefficient $\lambda$ controls the relative importance of the sparsity-inducing $L_1$ regularization term.  To apply the GL algorithm to our problem, we compute the covariance matrix between all pairs of regions for each cohort separately and directly apply the GL algorithm to learn a precision matrix for each cohort independently.  The GL algorithm cycles through the variables, fitting a modified lasso regression to each variable in turn.  The algorithm is fast, enabling problems with thousands of parameters to be solved efficiently.  In estimating functional connectivity across the entire brain, we can easily have thousands of parameters and small sample sizes, making this algorithm attractive. 
\\
\\         
\subsubsection{Joint Graphical Lasso}
\label{joint_glasso}
In many neuroimaging experiments we are searching for subtle changes in functional relationships between a subset of brain areas, i.e., we are not expecting the relationships between all brain regions to be different across cohorts.  In situations where we expect parameters to be shared across groups and have a small data set size relative to the number of parameters  being estimated, it has been shown by \citet{Danaher:2013fe}, using simulated data with a known amount of parameter sharing, that jointly estimated sparse precision matrices are closer to the true distributions than estimating them separately with the GL algorithm.  The learning problem is similar to equation (\ref{mle_glasso}) except we replace the $L_1$ norm penalty with a generic penalty function $P\left(\left\{\Phi\right\}\right)$ defined across the set of precision matrices $\left\{\Phi\right\}$:
 \begin{equation}
 \label{mle_joint}
\left[\left(\hat{\Sigma}^{(1)}\right)^{-1}, \hdots , \left(\hat{\Sigma}^{(G)}\right)^{-1} \right] =  \argmax_{\ \Phi \succ 0} \left( \sum^{G}_{g=1} n_g \left[ \log \det \Phi^{(g)} - \operatorname{tr}(S^{(g)}\Phi^{(g)}) \right] - P\left(\left\{\Phi\right\}\right) \right)
 \end{equation}
 where $\left[\left(\hat{\Sigma}^{(1)}\right)^{-1}, \hdots , \left(\hat{\Sigma}^{(G)}\right)^{-1} \right]$ are the estimates of the sparse precision matrices for all groups, $n_{g}$ is the number of observations in group $g$, and the first term inside the sum is the contribution from each group to the log-likelihood.  \citet{Danaher:2013fe} propose two convex penalty functions, the Fused Graphical Lasso (FGL) and Group Graphical Lasso (GGL) penalties.  FGL encourages groups to share both network structure and parameter values; the FGL penalty is given by:
 \begin{equation}
 \label{fgl}
P\left(\left\{\Phi\right\}\right) =  \lambda_{1} \sum_{g=1}^{G} \sum_{i \ne j} \left | \Phi_{i,j}^{(g)} \right | + \lambda_{2}  \sum_{g < g'} \sum_{i,j} \left |  \Phi_{i,j}^{(g)} - \Phi_{i,j}^{(g')} \right |
 \end{equation}
 where $\lambda_{1}$ and $\lambda_{2}$ are non-negative coefficients controlling the amount of sparsity and similarity in the precision matrices across groups respectively.  The first term in equation (\ref{fgl}) is similar to the $L_{1}$ penalty in graphical lasso, except that the sum is over the off-diagonal elements.  Increasing this penalty results in sparser networks.  The second term is the absolute difference in corresponding parameter estimates across the $G$ precision matrices that we are estimating, one for each cohort.  Increasing the $\lambda_{2}$ penalty will result in similar connectivity networks and strengths between pairs of variables across the groups.  Making $\lambda_{2}$ extremely large will result in identical networks across the groups.   
\\
\\   
In contrast, the GGL penalty encourages a shared pattern of sparsity, without requiring similarity between the parameter estimates across the groups.  The GGL objective is:
 \begin{equation}
 \label{ggl}
P\left(\left\{\Phi\right\}\right) =  \lambda_{1} \sum_{g=1}^{G} \sum_{i \ne j} \left | \Phi_{i,j}^{(g)} \right | + \lambda_{2}  \sum_{i \ne j} \left( \sum_{g=1}^{G} \Phi_{i,j}^{(g)^{2}} \right ) ^{\frac{1}{2}}
 \end{equation}
 where $\lambda_{1}$ and $\lambda_{2}$ are non-negative coefficients controlling the amount of sparsity and shared sparsity respectively.  The first term in equation (\ref{fgl}) is the same $L_{1}$ penalty as FGL; as before, making this penalty large results in sparser networks.  The second term encourages a similar pattern of sparsity in the precision matrices across the groups.  Interestingly, both penalty terms will contribute to sparsity resulting in a weaker form of network similarity than FGL, where patterns of sparsity (e.g., functional connectedness) are similar across the groups, but the strength of connections in the network are not penalized for being different. 
\\
\\
To estimate the networks using joint graphical lasso penalties, we use the alternating directions method of multipliers (ADMM) \cite{Boyd:2011bw}.  The algorithm is guaranteed to converge to the global optimum.  For details on applying ADMM to this problem see \citet{Danaher:2013fe}.  To apply the joint graphical lasso algorithms to our problem, we follow the same procedure as for graphical lasso except the precision matrices are learned jointly across cohorts.  Compared to graphical lasso, the joint lasso penalties have two coefficients to set, $\lambda_1$ and $\lambda_2$.  In section \ref{results} we will evaluate different ways of setting the penalties and whether the joint models improve accuracy in recovering the gold standard connectivity patterns and strengths across a range of data set sizes.    
   
 \subsection{Gold Standard Data}
 \label{gold_standard}
Ultimately we are interested in learning about differences in how the brain is sharing information across groups of subjects, evaluated with static molecular imaging techniques, because it may provide important information on overall brain function.  In practical settings, the amount of available data is small relative to the number of parameters needed to estimate whole-brain connectivity, which could result in both spurious connections and/or poor estimates of true connections due to dependencies in the parameters \cite{varoquaux2013learning}.  In section \ref {SICE} we described three models for functional connectivity that use regularization to, hopefully, improve predictive accuracy and reduce variance in our parameter estimates, at the expense of some bias, while also making the results more interpretable, focusing our attention on the strongest functional connections \cite{Tibshirani:1996bc}.  In order to evaluate the performance of the inverse covariance models in finding true functional connections in molecular imaging data, we require a large data set with many samples.  
\\
\\
The Amen Clinics Inc. (\url{http://www.amenclinics.com/}) has been collecting technetium-99m hexamethylpropyleneamine oxide (Tc99m HMPAO) SPECT scans for many years on a variety of disorders such as attention deficit hyperactivity disorder, depression, anxiety, and behavioral problems, among others.   The Amen Clinics Inc. provided us with a SPECT data set to use as a gold standard, consisting of 11,906 males (avg. age 29.2 $\pm$ 17.6) and 7,550 females (avg. age 35.6 $\pm$ 18.4)  \cite{amen2015}.  This is the largest SPECT data set that we know of.  On average, 55\% of the subjects have mood disorders, 7\% bipolar, 25\% depression, 47\% ADHD, 45\% anxiety, 10\% substance abuse, and 32\% brain trauma.  Although the data set consists of a variety of disorders and comorbitidies, for this purpose we need two or more large cohorts that have both shared and unshared functional connections, to evaluate how well the three inverse covariance models recover these patterns as a function of sample size.  
\\
\\
\subsubsection{Regions of Interest}
\label{ROI}
Prior to performing an analysis of functional connectivity, we must define the nodes of our network.  In our gold standard dataset, structural scans were unavailable and thus a popular choice is to use a standard atlas to define regions of interest (ROIs) such as the AAL atlas from \citet{TzourioMazoyer:2002bi} and average the functional voxels contained in each anatomical boundary.  Unfortunately, this approach essentially ignores the functional data in defining the ROIs and is prone to decreased sensitivity by averaging signals from a small number of activated voxels with noise \cite{Poldrack:2006dj}. Here we take a pragmatic approach, given the data we have available for evaluating the inverse covariance models, and use Gaussian finite mixture modeling to find functional clusters constrained by anatomically-defined boundaries for use as nodes in the network.   Our motivation is to find a set of regions that are consistent with the observed functional data across all subjects, while also allowing the nodes to be interpreted with respect to an anatomical reference.   Details about applying Gaussian finite mixture modeling to define regions of interest can be found in \ref{clustering}.  Briefly, for each anatomical region defined in the AAL atlas, we use the observed counts for each voxel contained in the anatomical region, across all subjects, and perform Gaussian finite mixture modeling to learn clusters of functional activity.  We then use the Bayesian information criterion (BIC) score to select the number of clusters (i.e., $K$) for each region that best fits our observed data \cite{Schwarz:1978kf}.  Once the number of clusters have been selected for each anatomical region, we compute values for each cluster by averaging over the number of detections at each voxel, weighted by their cluster membership probabilities, for each subject separately.  We use these cluster averages, computed for each subject, in the sparse inverse covariance models.        
\\
\\     
\subsubsection{Scan Preparation}

Subjects in the Amen data set were injected with an age/weight appropriate dose of Tc99m HMPAO and were at rest during uptake.  All subjects were scanned on a high-resolution Picker Prism 3000 triple-headed gamma camera with fan beam collimators.  The original reconstructed image matrices were 128x128x29 voxels with sizes of 2.16mm x 2.16mm x 6.48mm and values representing counts.  The images were spatially normalized to the MNI atlas using SPM8 software \cite{Penny:2011ve}, resulting in image matrices of 79 x 95 x 68 voxels in x, y, and z dimensions respectively with isotropic 2mm voxel sizes.  The Automated Anatomical Labeling (AAL) atlas was used to define the brain regions based on the anatomical parcellations available in the atlas because no structural imaging data was available \cite{TzourioMazoyer:2002bi}.  
\\
\\
After normalization, the clustering model (see \ref{clustering}) for $K\in\left\{1,\dotsc , 25\right\}$ was run across the entire data set, for each region in the AAL atlas, except the cerebellum.  The cerebellum was not included due to missing data in the lower slices of some scans.  The clustering model, after selecting the best model for each anatomical region using the BIC scores, resulted in 180 clusters across the brain.  The cluster averages for each subject, stratified by cohorts, were computed and mean centered.   Although the sample is large relative to the number of variables, there is still measurement noise in the data.   To remove some of this noise, we run the graphical lasso algorithm with light regularization ($\lambda=0.1$), resulting in precision matrices with 2119 pairwise connected regions in males (partial correlation range -0.66-0.17), 2130 pairwise connections in females (partial correlation range -0.64-0.17), and 1750 shared connections ($\approx 82\%$) across both cohorts.  These precision matrices will be used in section \ref{results} as a gold standard, to quantitatively evaluate the performance of the inverse covariance estimation models in recovering the patterns as a function of the amount of available training data.

%\begin{figure}
%\centering
 %\includegraphics[height=3.5in]{gs_clusters.png}
 %\caption[Gold standard SPECT cluster results] {Results from the clustering model run on the gold standard SPECT data set (n=19,456) collapsed across groups.  Ellipsoids show the  $\pm 1$ standard deviation region around each cluster mean. }
% \label{gs_clustering}
 %\end{figure}

%\begin{figure}
%\centering
%\begin{subfigure}[Males] {0.49\textwidth}
 %\includegraphics[width=\textwidth]{GoldStandardEmpiricalCorrMatrix_Grp1.png}
 %\caption [Gold standard males empirical correlation matrix]{Males}
 %\label{fig:Males}
 %\end{subfigure}
%\begin{subfigure}[Females] {0.49\textwidth}
 %\includegraphics[width=\textwidth]{GoldStandardEmpiricalCorrMatrix_Grp2.png}
 %\caption [Gold standard females empirical correlation matrix]{Females}
 %\label{fig:Females}
 %\end{subfigure}
 %\caption[Gold standard male and female empirical correlation matrices] {Empirical correlation matrices computed from the average values for each cluster for (a) males (n=11,906) and (b) females (n=7,550).  Correlations range from -0.5 to 1.0}
 %\label{gs_emp_corr_mat}
 %\end{figure}

\subsection{Model Summary}
\begin{figure}
\centering
 \includegraphics[width=5.0in]{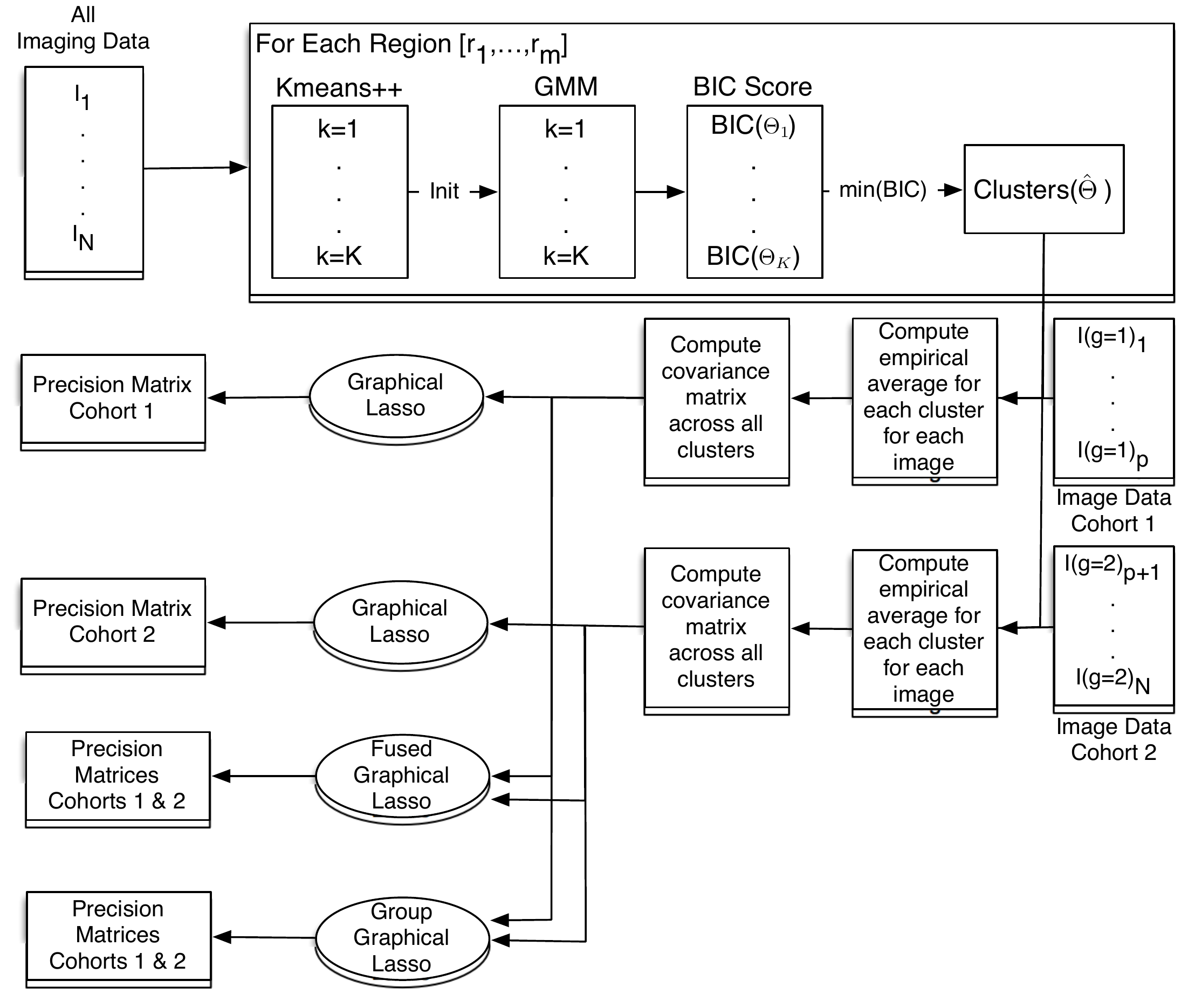}
 \caption [Model flow chart]{Summary of the methods for learning functional clusters and precision matrices using graphical lasso, fused graphical lasso, and group graphical lasso algorithms in group-based static molecular imaging data.  Note, in this example there are only two cohorts but the process is the same for additional groups.}
 \label{model_flowchart}
 \end{figure}

\begin{figure}
\centering
 \includegraphics[width=6.0in]{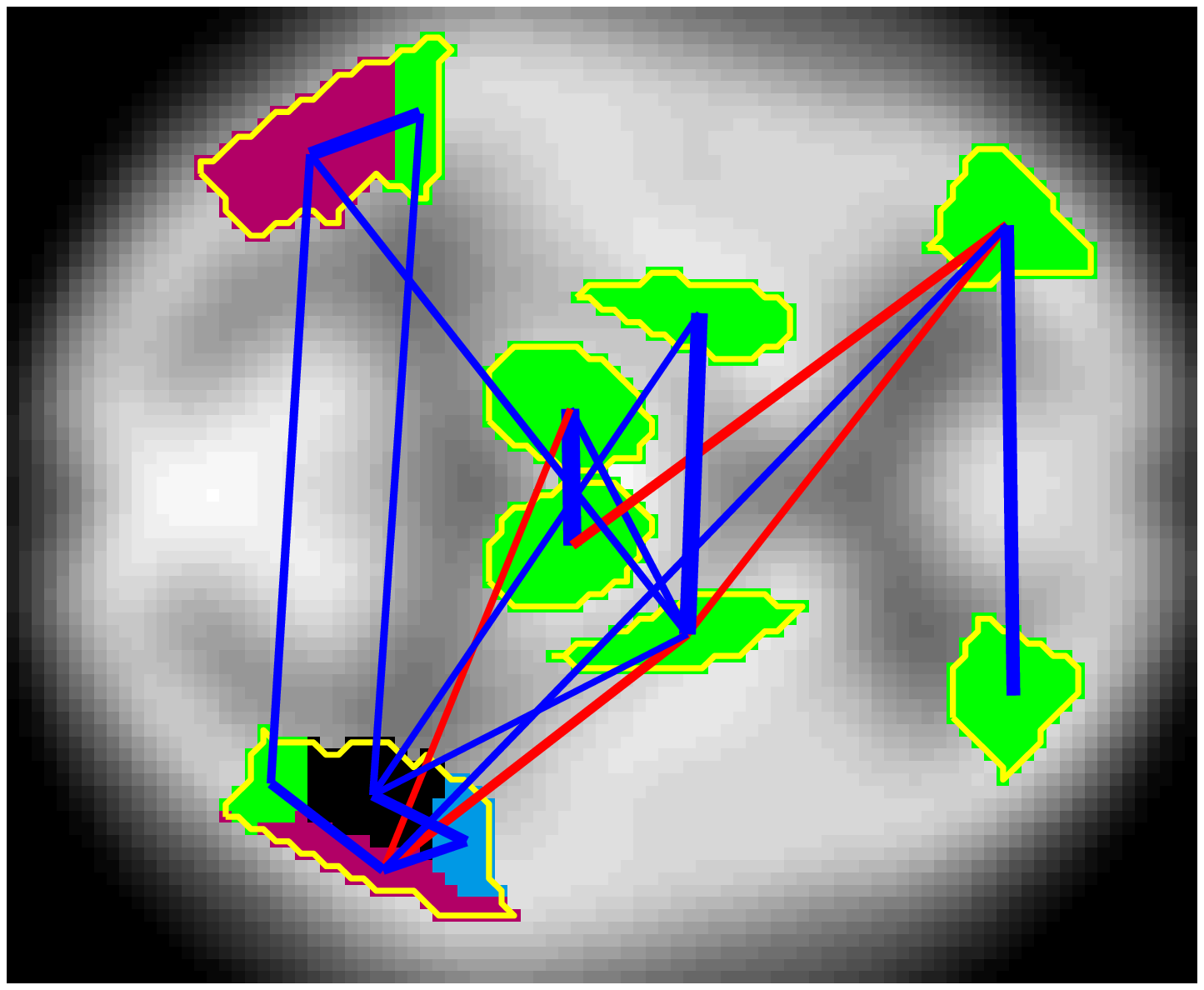}
 \caption [Example clustering and inverse covariance estimation]{Example of clustering and inverse covariance modeling in a two-dimensional SPECT scan slice.  Yellow outlines correspond to anatomical regions from an atlas, voxels are colored by their cluster membership within each anatomical region, and lines connecting clusters correspond to the non-zero parameters in the inverse covariance matrix between pairs of clusters and are scaled by the magnitude of the partial correlations, where blue edges indicate negative partial correlations and red areas positive partial correlations.}
 \label{cluster_example}
 \end{figure}

The clustering and sparse inverse covariance algorithms are summarized graphically in Figure \ref{model_flowchart}.  The  entire set of imaging data is collapsed across cohorts prior to clustering.  For each region, kmeans++ is run for each setting of $K$ and used to initialize a Gaussian mixture model \cite{Arthur:2007tv}.  The resulting clusters are evaluated using the BIC scores and the solution with the minimum BIC score for each region are selected and used as regions of interest (see \ref{clustering}).  The region of interest averages are then calculated for each subject and each cluster across all regions.  For each cohort, we then compute the empirical covariance matrix among all pairs of clusters and learn precision matrices using the graphical lasso, fused, and group graphical lasso algorithms.  This process results in inverse covariance matrices with varying amounts of non-zero entries which are depicted in a graph structure by adding an edge between pairs of variables.  Figure \ref{cluster_example}  shows an example of clustering and the graphical lasso model run with eight anatomical regions (i.e., yellow region outlines) where voxels have been colored by their cluster membership within each anatomical region.  The edges in the graph indicate non-zero entries in the inverse covariance matrix between pairs of clusters and are scaled by the magnitude of the partial correlations, where blue edges indicate negative partial correlations and red areas positive partial correlations.   Pairs of clusters that are not connected in the figure have zero entries in the inverse covariance matrix, reflecting their conditional independence.  Note, in the experiments discussed in section \ref{results}, these procedures where run in three-dimensions using the whole brain.        

\section{Results}
\label{results}
In this section we provide a quantitative evaluation of the sparse inverse covariance models in terms of correct connections and connection strengths, across different sample sizes compared to the gold standard precision matrices described in section \ref{gold_standard}.  The intent of our experiments are to evaluate the performance of the inverse covariance models at sample sizes typically used in functional imaging studies.  Each of the models described in section \ref{SICE} requires a setting for the amount of regularization, which affects the sparsity of the results and the overall network structure; therefore, we evaluate two methods of setting the regularization weights.  In section \ref{gt_results} we use the gold standard precision matrices directly to determine optimal settings for the $\lambda_2$ regularization coefficients in the FGL and GGL models and compare receiver operating characteristic (ROC) curves of true-positive (TP) and false-positive (FP) connections across all models by varying $\lambda$ in GL and $\lambda_1$ in FGL and GGL.  In practice, one does not typically have access to a gold standard.  In section \ref{ct_results} we use cross-validation to determine the regularization weights and compare the resulting precision matrices with those from the gold standard.

\subsubsection{Using Gold Standard for Parameter Settings}
\label{gt_results}
\begin{figure}
\centering
%\begin{subfigure}[n=5] {0.55\textwidth}
%\includegraphics[width=\textwidth]{TPFP_Edge_vs_GS_5.eps}
% \caption {n=5}
% \label{fig:n=5}
% \end{subfigure}
\begin{subfigure}[n=10] {0.45\textwidth}
 \includegraphics[width=\textwidth]{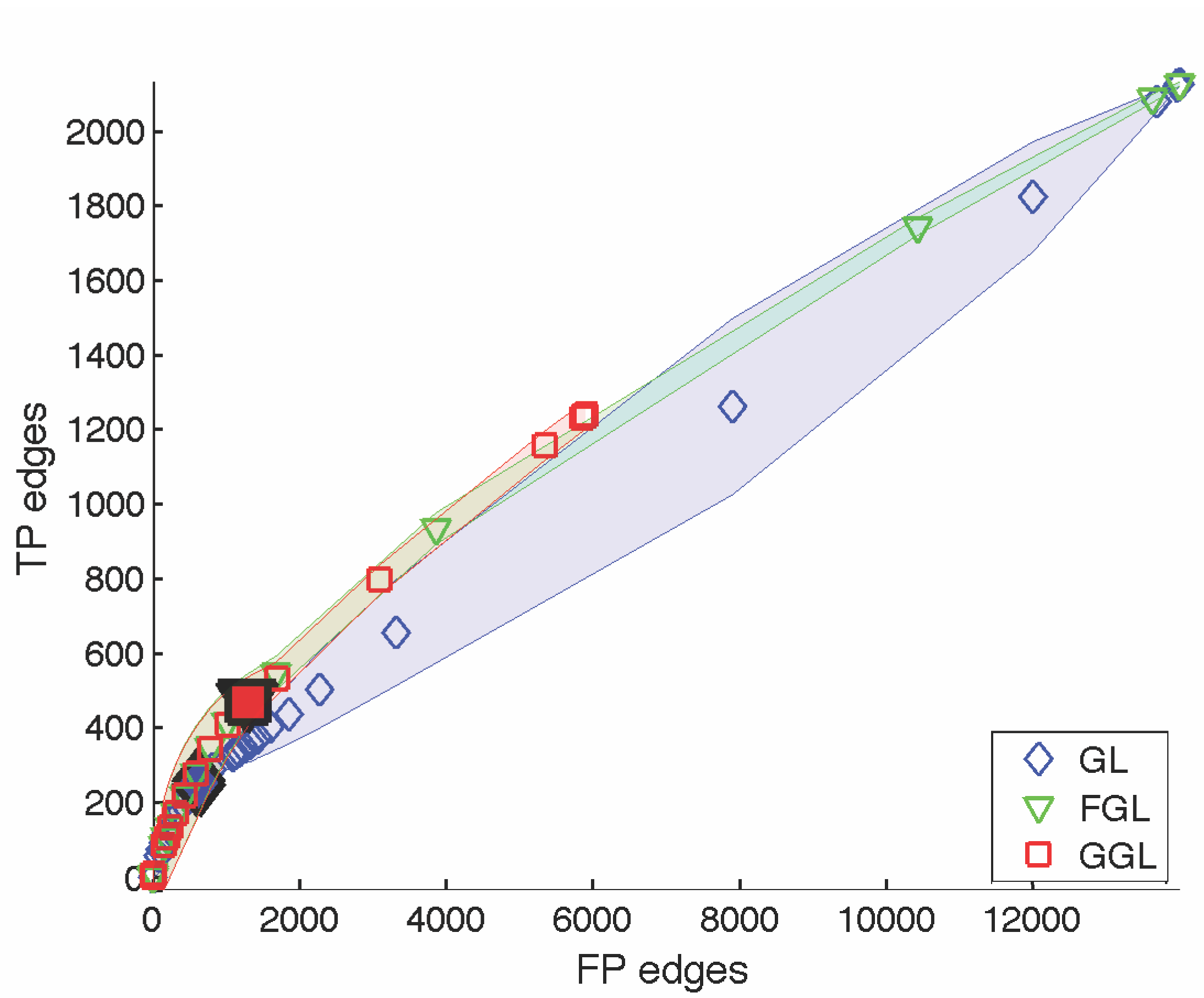}
 \caption {n=10}
 \label{fig:n=10}
 \end{subfigure}
\begin{subfigure}[n=25] {0.45\textwidth}
 \includegraphics[width=\textwidth]{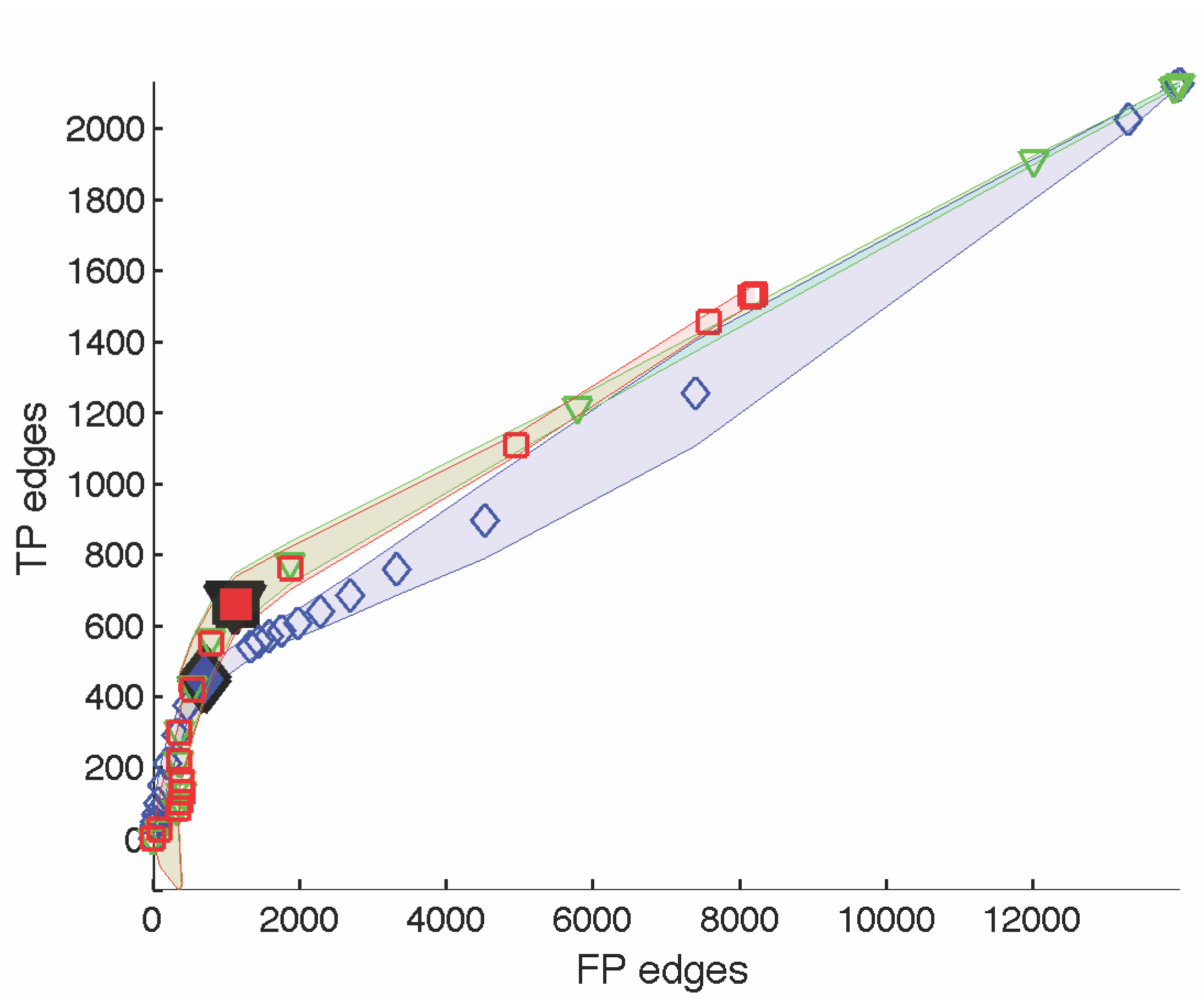}
 \caption {n=25}
 \label{fig:n=25}
 \end{subfigure}
 \begin{subfigure}[n=50] {0.45\textwidth}
 \includegraphics[width=\textwidth]{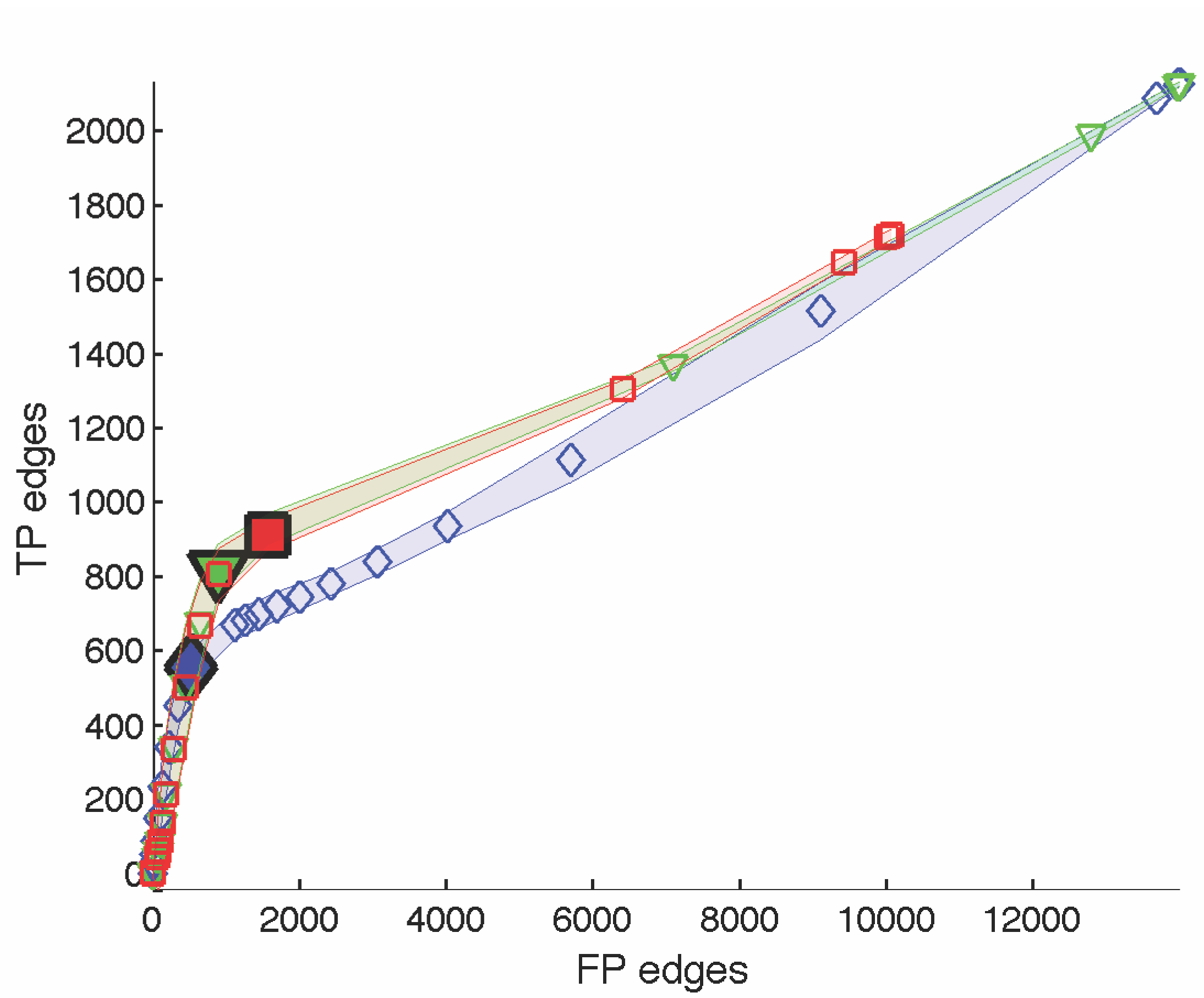}
 \caption {n=50}
 \label{fig:n=50}
 \end{subfigure}
 \begin{subfigure}[n=100] {0.45\textwidth}
 \includegraphics[width=\textwidth]{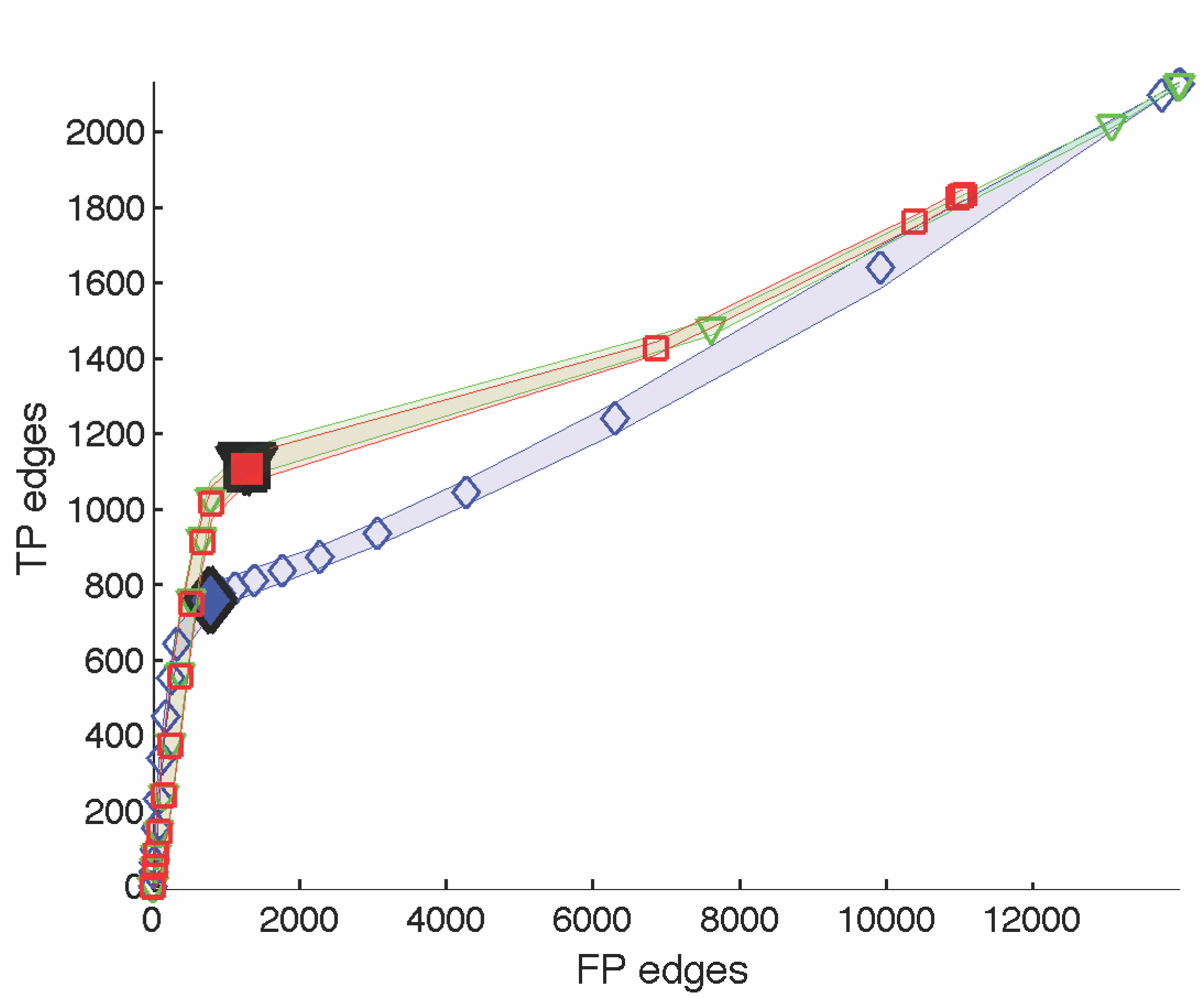}
 \caption {n=100}
 \label{fig:n=100}
 \end{subfigure}
    \caption [TP/FP for edge selection vs. gold standard $n \in \{10,25,50,100\}$]{True-positive (TP) vs. false-positive (FP) edges between the test precision matrices and the gold standard for sample sizes $n \in \{10,25,50,100\}$ using GL, FGL, and GGL inverse covariance models.  Small markers indicate the means across the 10 test subsets and both cohorts.  Colored regions show the unit standard deviation about the means. Large markers identify points with lowest average SSE compared to gold standard.}
 \label{TPFPedges1}
 \end{figure}

\begin{figure}
\centering
 \begin{subfigure}[n=250] {0.45\textwidth}
 \includegraphics[width=\textwidth]{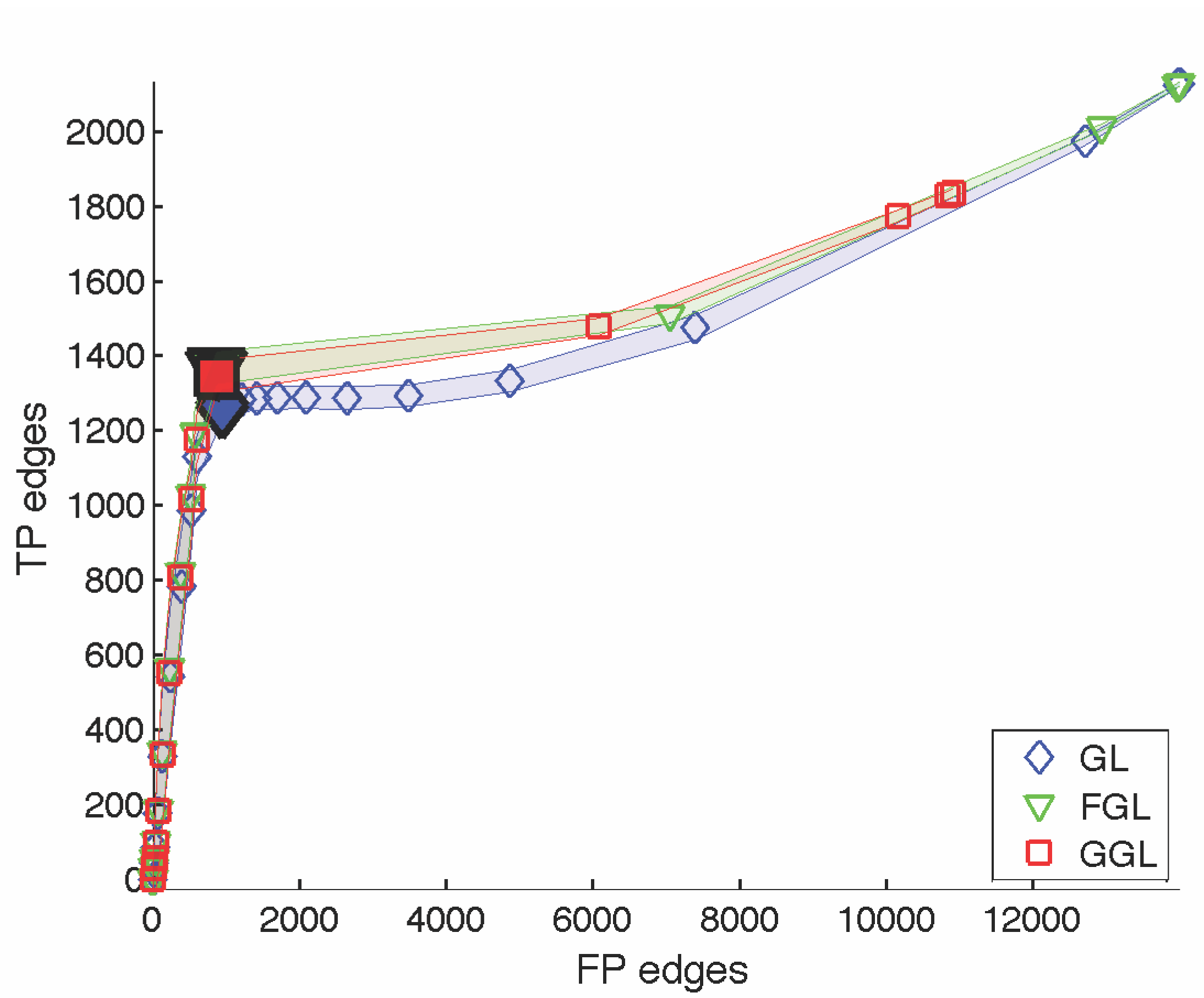}
 \caption {n=250}
 \label{fig:n=250}
 \end{subfigure}
 \begin{subfigure}[n=500] {0.45\textwidth}
 \includegraphics[width=\textwidth]{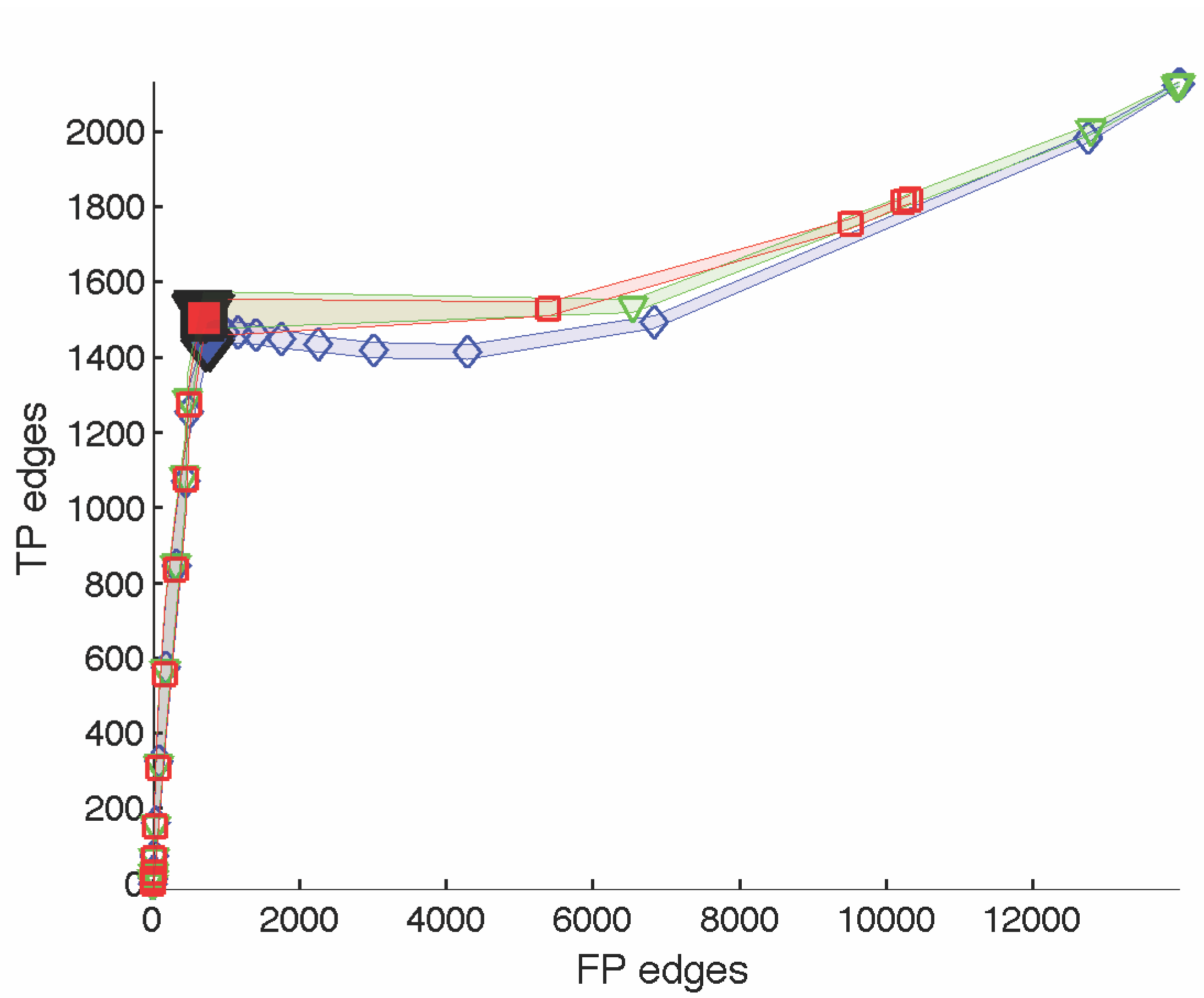}
 \caption {n=500}
 \label{fig:n=500}
 \end{subfigure}
 \begin{subfigure}[n=750] {0.45\textwidth}
 \includegraphics[width=\textwidth]{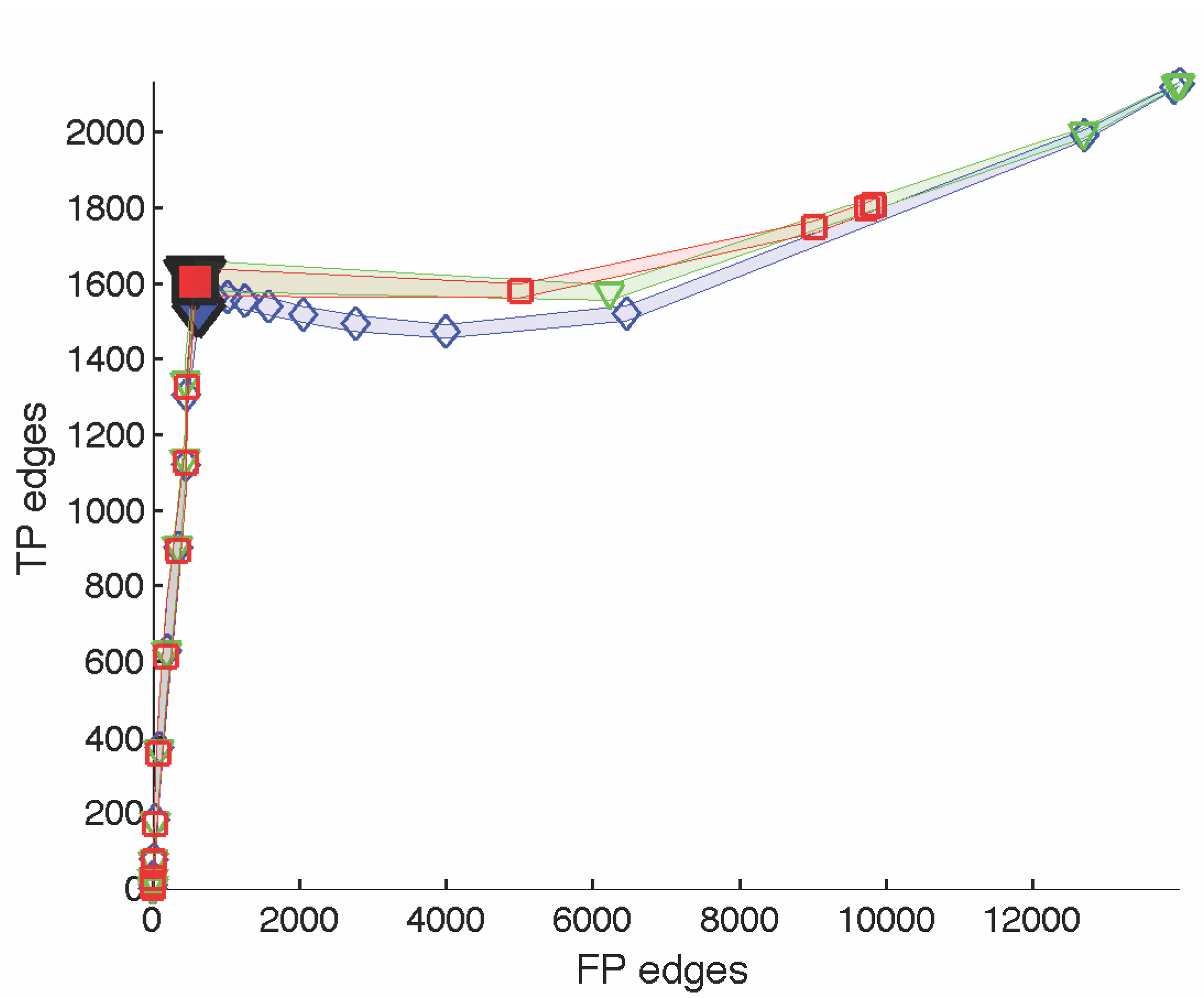}
 \caption {n=750}
 \label{fig:n=750}
 \end{subfigure}
 \begin{subfigure}[n=1000] {0.45\textwidth}
 \includegraphics[width=\textwidth]{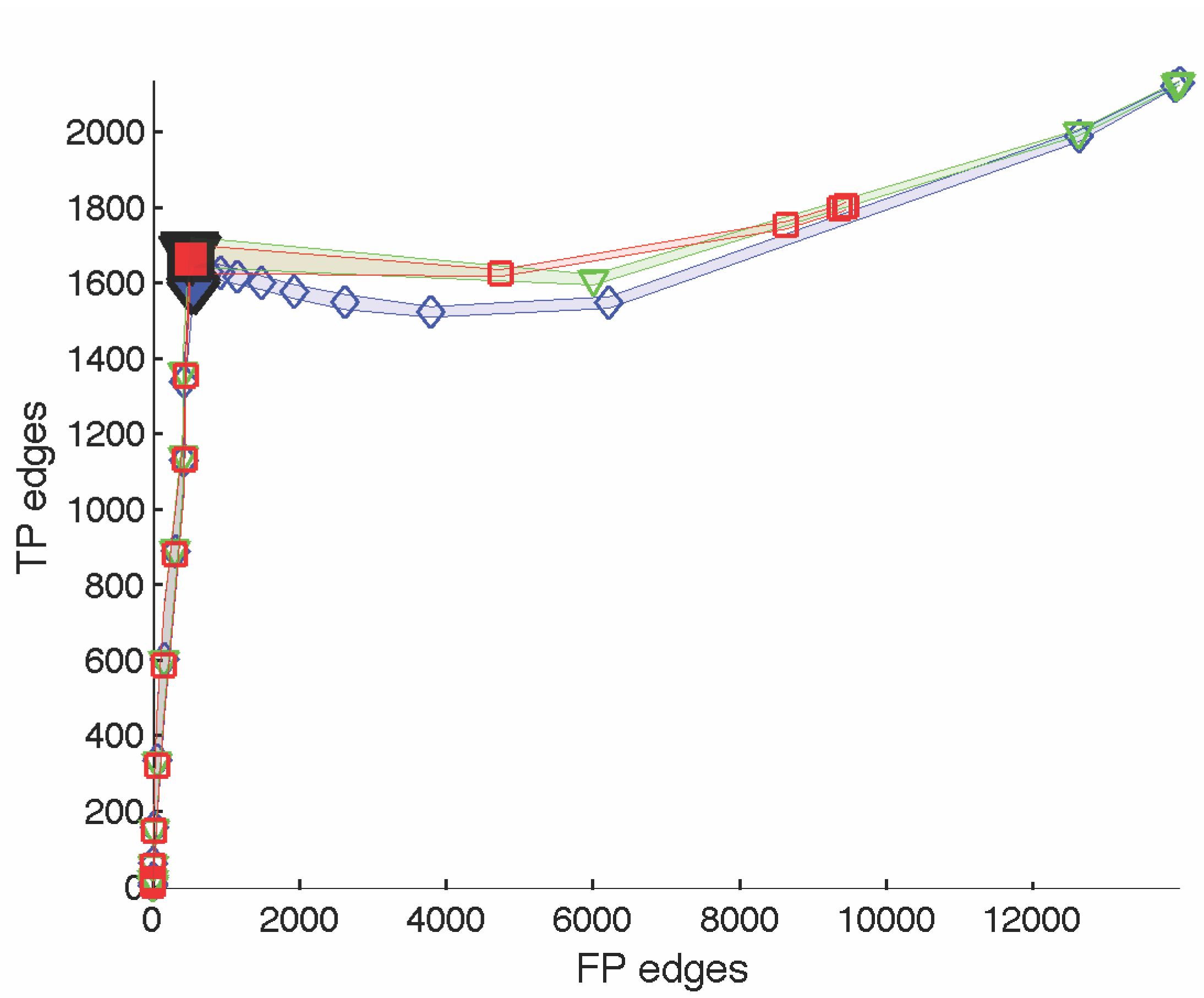}
 \caption {n=1000}
 \label{fig:n=1000}
 \end{subfigure}
 %%\begin{subfigure}[n=1000] {0.49\textwidth}
 %% \includegraphics[width=\textwidth]{TPFPEdge_vs_GS_1000.png}
 %% \caption {n=1000}
 %% \label{fig:n=1000}
 %% \end{subfigure}
 %%\begin{subfigure}[n=1500] {0.49\textwidth}
 %%\includegraphics[width=\textwidth]{TPFPEdge_vs_GS_1500.png}
 %% \caption {n=1500}
 %%\label{fig:n=1500}
 %%\end{subfigure}
 %%  \begin{subfigure}[n=2000] {0.49\textwidth}
 %% \includegraphics[width=\textwidth]{TPFPEdge_vs_GS_2000.png}
 %% \caption {n=2000}
 %% \label{fig:n=2000}
 %% \end{subfigure}
 %%  \begin{subfigure}[n=2500] {0.49\textwidth}
 %% \includegraphics[width=\textwidth]{TPFPEdge_vs_GS_2500.png}
 %% \caption {n=2500}
 %% \label{fig:n=2500}
 %% \end{subfigure}
  \caption[TP/FP for edge selection vs. gold standard $n \in \{250,500,750,1000\}$] {True-positive (TP) vs. false-positive (FP) edges between the test precision matrices and the gold standard for sample sizes $n \in \{250,500,750,1000\}$ using GL, FGL, and GGL inverse covariance models.  Small markers indicate the means across the 10 test subsets and both cohorts.  Colored regions show the unit standard deviation about the means. Large markers identify points with lowest average SSE compared to gold standard.}
 \label{TPFPedges2}
 \end{figure}

\begin{figure}
\centering
\begin{subfigure}[n=1500] {0.45\textwidth}
 \includegraphics[width=\textwidth]{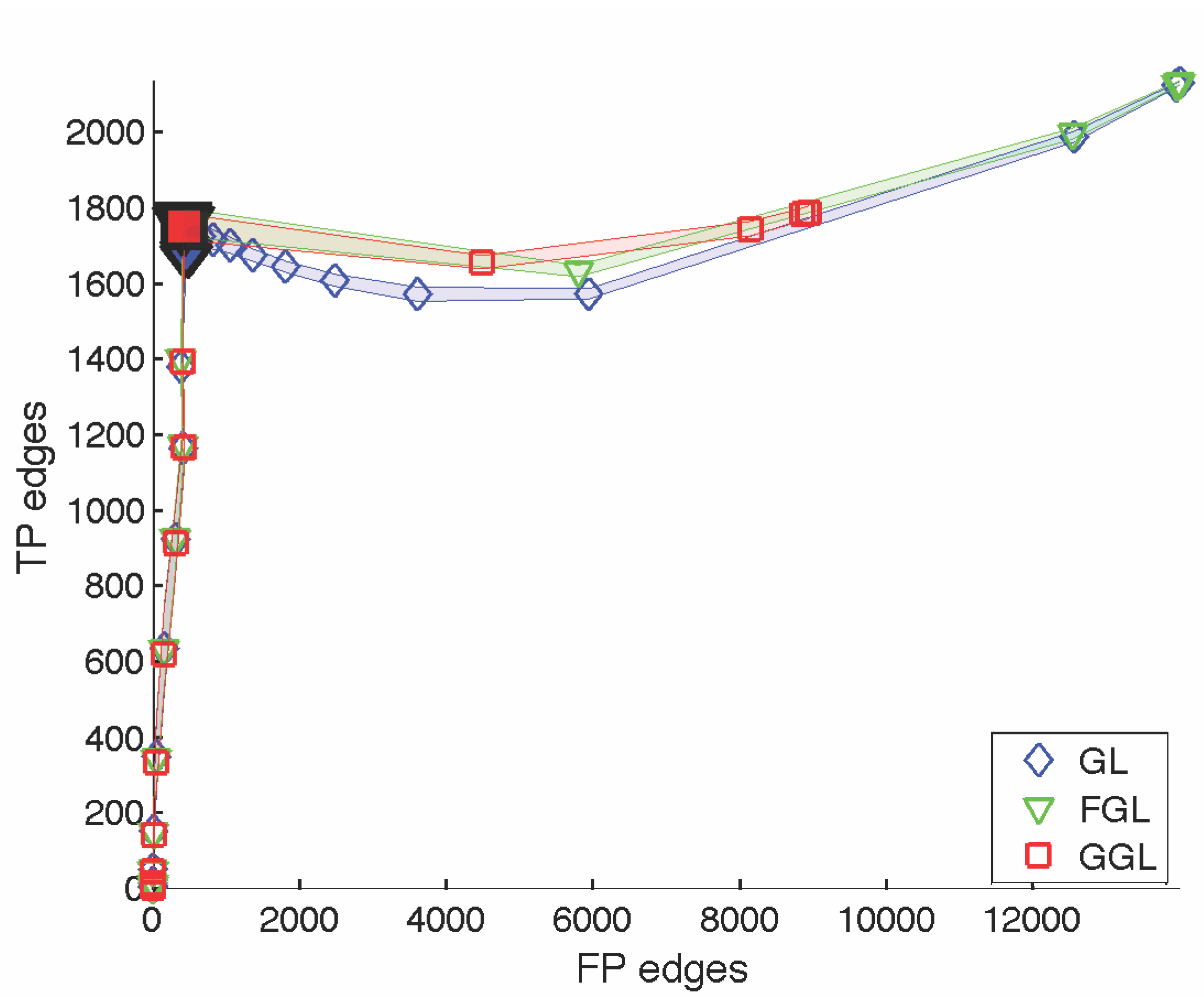}
 \caption {n=1500}
 \label{fig:n=1500}
 \end{subfigure}
\begin{subfigure}[n=2000] {0.45\textwidth}
 \includegraphics[width=\textwidth]{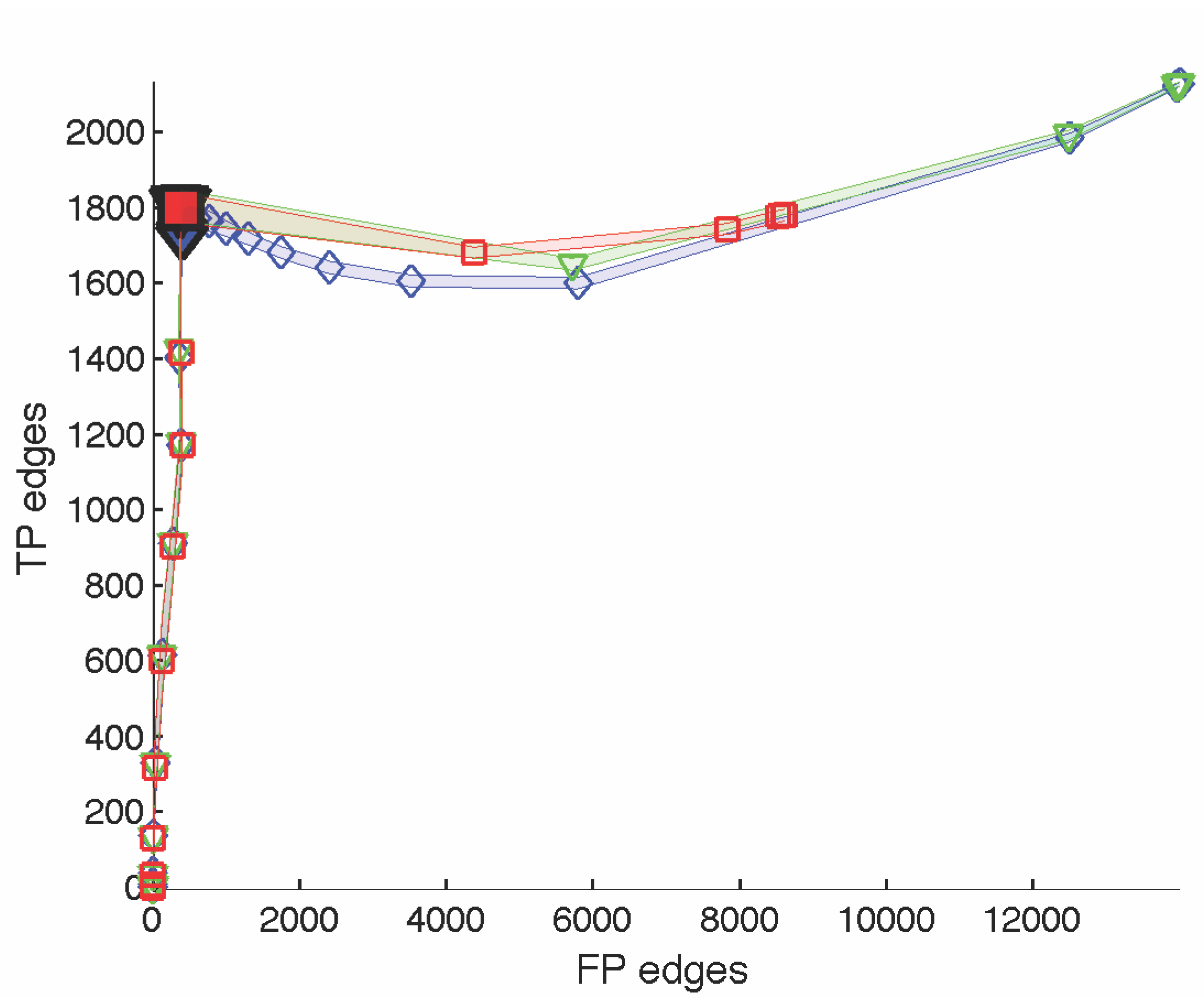}
 \caption {n=2000}
 \label{fig:n=2000}
 \end{subfigure}
 \begin{subfigure}[n=2500] {0.45\textwidth}
 \includegraphics[width=\textwidth]{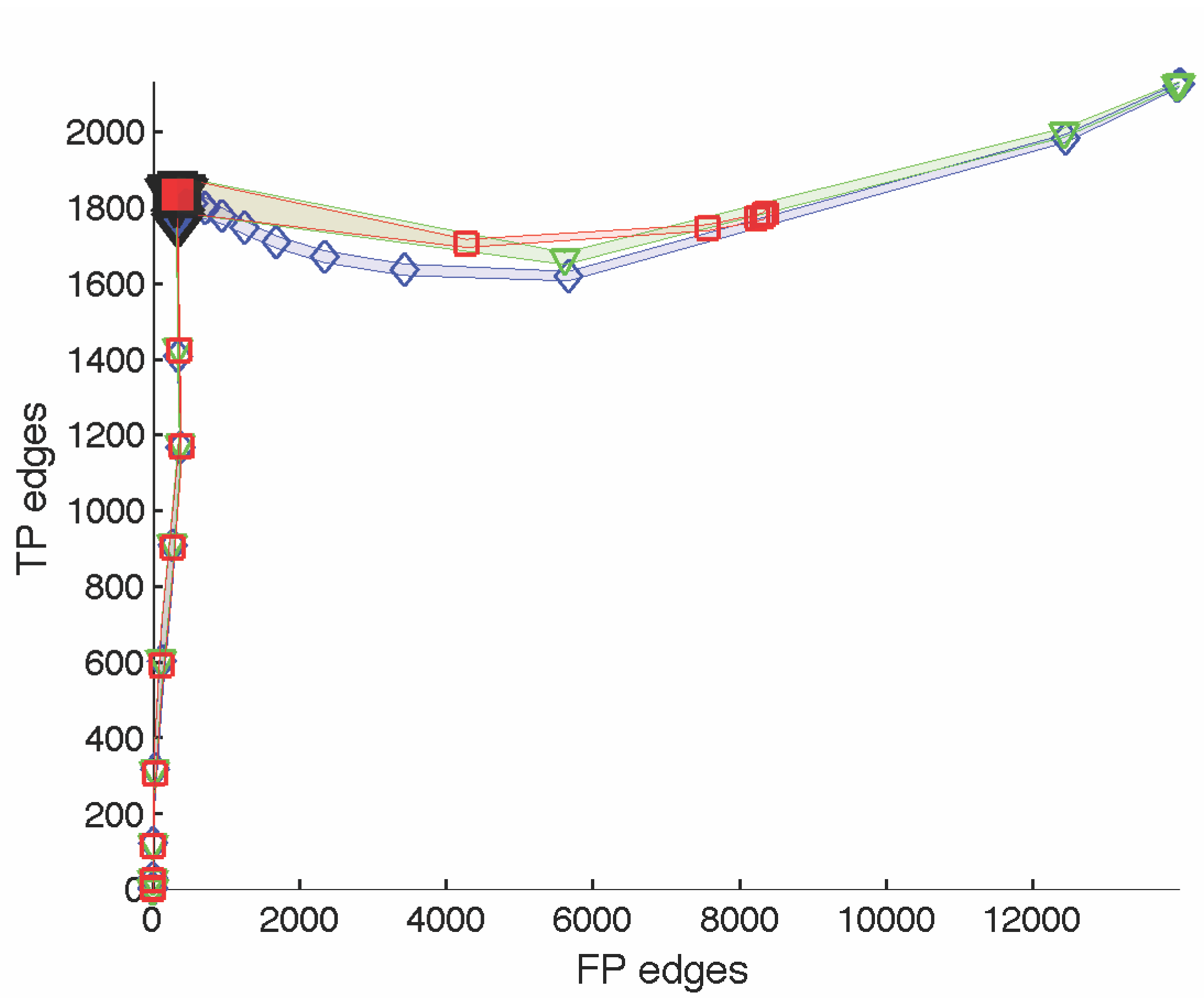}
 \caption {n=2500}
 \label{fig:n=2500}
 \end{subfigure}
    \caption [TP/FP for edge selection vs. gold standard $n \in \{1500,2000,2500\}$]{True-positive (TP) vs. false-positive (FP) edges between the test precision matrices and the gold standard for sample sizes $n \in \{1500,2000,2500\}$ using GL, FGL, and GGL inverse covariance models.  Small markers indicate the means across the 10 test subsets and both cohorts.  Colored regions show the unit standard deviation about the means. Large markers identify points with lowest average SSE compared to gold standard.}
 \label{TPFPedges3}
 \end{figure}

In this experiment we evaluate the sparse inverse covariance models by sample size, using the gold standard precision matrices to determine settings for the regularization parameters.  To compare GL, FGL, and GGL models, we follow a procedure similar to \citet{Danaher:2013fe}.  We fix the regularization coefficients of the group penalties (i.e., $\lambda_2$) and vary the $\lambda_1$ penalty in the FGL and GGL models and the $\lambda$ penalty in the GL model, creating ROC curves of model accuracy in predicting correct functional connections.  Further, we compare the strength of connections relative to the gold standard precision matrices.  To determine fixed settings for the group penalties, we use the following procedure.  We first draw two random samples (n=500) from the gold standard data set, stratified by group.  We then compute the cluster averages using the clusters found with the gold standard data set described in section \ref{gold_standard}.  Using these data, we perform a grid search over the $\lambda_1$ and $\lambda_2$ parameter space, for the FGL and GGL models separately, comparing the learned precision matrices ($\hat{\Phi}$) to those of the gold standard ($\Phi^{*}$) in terms of their $L_{1}$ difference $\left ( \sum^{G}_{g=1}  \left |\hat{\Phi}_{g} -  \Phi^{*}_{g} \right | \right )$.  We select the settings that result in the minimum $L_1$ difference from the gold standard.  For the FGL model, the minimum was found with settings $\lambda_1=0.01$ and $\lambda_2=0.1$.  For the GGL model, the minimum was found with settings $\lambda_1=0.001$ and $\lambda_2=0.01$.  We will use these fixed settings for $\lambda_2$ in our comparison of the group regularized models (i.e., FGL and GGL) with GL.   
\\
\\
Next, we draw 10 subsets of images, with replacement, stratified by groups, with equal sizes \\ $N \in \{10,25,50,100,250,500,750,1000,1500,2000,2500\}$ from the gold standard data set, after removing the subjects used in the parameter searching routine described above.  Using the clusters found for the gold standard data set described in section \ref{gold_standard}, we compute the cluster averages and mean center the data.  Next, we learn precision matrices using the GL model across a wide range of $\lambda$ regularization penalties.  Similarly, using the fixed $\lambda_2$ settings, we learn precision matrices using the FGL and GGL models across the same range of $\lambda_1$ regularization settings as with GL.  
\\
\\
The results showing the average number (across both groups and 10 subsets) of true-positive edges (TP) correctly identified (i.e., non-zero entries in the test precision matrices match those from the gold standard), compared to the average number of false-positive edges (FP) incorrectly identified  (i.e., non-zero entries in the test precision matrices unmatched with the gold standard) by sample sizes are shown in Figures \ref{TPFPedges1}-\ref{TPFPedges3}.  The colored regions around each marker indicate the unit standard deviation region about the means.  The large markers identify points on the curves with the lowest sum of squared errors (SSE) between the test precision matrices and the gold standard for each model, indicating the point on the curves where connection strengths in the test precision matrices best match the gold standard.  Interestingly, these occur in regions with a high proportion of true-positive edges relative to false positive edges.  Visually, the FGL and GGL model curves generally dominate the GL curves for most sample sizes.   The biggest difference between the joint models and graphical lasso are in the moderate sample sizes $N \in \{25,50,100\}$, where we see the largest benefit from parameter sharing across the groups.  Both joint models, in general, perform equally well in terms of the presence or absence of edges.   The average areas under the TP/FP edge curves (AUC) across the subsets, by sample sizes, are shown in Table \ref{auc_ttest} (columns 2-4).  The T-test and Bonferroni corrected p-values, comparing all combinations of the models by sample sizes, are shown in columns 5-7.   The AUCs are statistically lower in the GL model (i.e., negative t-scores) than both the FGL and GGL models at data set sizes less than or equal to 1000, with the moderate sample sizes showing the largest decreases.  At data set sizes above 2000, the GL model outperforms the joint models, which is related to having enough data to accurately estimate the precision matrices without the added benefit of parameter sharing in the joint models, and the fact that we used the GL algorithm to remove noise when creating the gold standard precision matrices.  
\\
\\
To evaluate how well the entries in the precision matrices corresponding to the strength of functional connections match those of the gold standard data set, we calculate the SSE between the test precision matrices and the gold standard by model and sample size.  The average minimum SSEs (across both groups and 10 subsets) are shown in Table \ref{sse_ttest} (columns 2-4).  The T-test and Bonferroni corrected p-values comparing all combinations of the models by sample sizes are shown in Table \ref{sse_ttest} (columns 5-7).  The results are similar to the TP/FP edge results, where the joint models perform better than the GL model at moderate sample sizes, with the largest difference at $N=100$.  Between the FGL and GGL joint models, the FGL model performs better, yielding lower SSEs compared to the gold standard.    Unlike the TP/FP edge results, the FGL model achieves a lower SSE than the GL model, even for the larger data sets (i.e $N\in\{1500,2000,2500\}$), whereas GGL does not.  To evaluate the effect of the clustering model on these results, we repeated our experiments by averaging all the voxels within each region of interest instead of using the clustering model.  The results from the three sparse inverse covariance models are shown in Appendix Tables \ref{auc_ttest_nocluster} and \ref{sse_ttest_nocluster} and are consistent with those presented in this section.

%\begin{landscape}


%\end{landscape}

\begin{table}[ht]
\caption[TP/FP edge area under ROC curve]{TP/FP Edge Area Under Curve (AUC) T-test Comparison by Model and Sample Size using Gold Standard for Parameter Settings}
\centering
\resizebox{\columnwidth}{!} {%
\begin{tabular}{c c c c c c c }
\hline\hline
 & GL  &  FGL   & GGL  &GL-FGL  & GL-GGL & FGL-GGL \\
\hline
N & $\overline{AUC}\pm std$ & $\overline{AUC}\pm std$ & $\overline{AUC}\pm std$ & T(P Bonferroni) & T(P Bonferroni) & T(P Bonferroni) \\
\hline\hline
2500 &	0.820$\pm$0.003 & 0.790$\pm$0.008 &	0.810$\pm$0.008 & 10.74(\textbf{1.15E-07}) & 3.93(\textbf{3.79E-02}) & -5.51(\textbf{1.22E-03}) \\
\hline
2000 &	0.812$\pm$0.004 & 0.797$\pm$0.015	& 0.806$\pm$0.013 & 2.88(3.87E-01) &	1.31(2.08E-01) & -1.33(1.00e00) \\
\hline
1500 & 	0.799$\pm$0.004 &	0.792$\pm$0.018 &	0.806$\pm$0.013 & 1.19(1.00e00) &  -1.44(1.00e00) &	-1.87(1.00e00) \\
\hline
1000 &	0.784$\pm$0.004 & 0.799$\pm$0.005 &	0.811$\pm$0.003 	& -8.14(\textbf{7.52E-06}) &	-18.18(\textbf{1.93E-11}) &	-6.70(\textbf{1.09E-04}) \\
\hline
750 &	0.769$\pm$0.004 	& 0.789$\pm$0.007 &	0.797$\pm$0.005 	& -8.26(\textbf{6.07E-06}) &	-14.16(\textbf{1.31E-09}) &	-2.79(4.68E-01) \\
\hline
500	& 0.683$\pm$0.006 & 	0.776$\pm$0.006 &	0.779$\pm$0.004 &	-36.99(\textbf{7.65E-17}) &	-45.47(\textbf{1.93E-18}) & 	-1.27(1.00e00) \\
\hline
250	& 0.660$\pm$0.003 &	 0.746$\pm$0.007  &	0.745$\pm$0.007 &	-35.21(\textbf{1.83E-16}) & 	-35.30(\textbf{1.75E-16}) &		0.13(1.00e00) \\
\hline
100	& 0.631$\pm$0.006 &	 0.696$\pm$0.006 &	0.695$\pm$0.006 &	-24.07(\textbf{1.50E-13}) &	-23.26(\textbf{2.73E-13}) &		0.21(1.00e00) \\
\hline
50 & 0.604$\pm$0.004 &	0.652$\pm$0.005 &	0.653$\pm$0.005 &	-22.95(\textbf{3.44E-13}) &	-24.43(\textbf{1.16E-13}) &	-0.45(1.00e00) \\
\hline
25 & 	0.572$\pm$0.004 &	0.621$\pm$0.013 &	0.626$\pm$0.012 &	-11.28(\textbf{5.28E-08})	 & -13.19(\textbf{4.26E-09})	& 	-0.85(1.00e00) \\
\hline
10 &	0.537$\pm$0.007 &	0.601$\pm$0.013 &	0.604$\pm$0.013 &	-13.24(\textbf{3.98E-09}) &	-14.30(\textbf{1.12E-09}) &	-0.58(1.00e00) \\
\hline\hline
\end{tabular}%
}
\label{auc_ttest}
\end{table}
\begin{table}[ht]
\caption[Sum of squared errors with gold standard by model and sample size]{Sum of Squared Errors (SSE) with Gold Standard by Model and Sample Size}
\centering
\resizebox{\columnwidth}{!} {%
\begin{tabular}{c c c c c c c}
\hline\hline
& GL  &  FGL   & GGL &GL-FGL  & GL-GGL & FGL-GGL \\
\hline
N & $\overline{SSE}\pm std$ & $\overline{SSE}\pm std$ & $\overline{SSE}\pm std$ & T(P Bonferroni) & T(P Bonferroni) & T(P Bonferroni) \\
\hline\hline
2500 & 1.35$\pm$0.27 & 1.34$\pm$0.18 & 1.74$\pm$0.16 & 0.11(1.00E+00)	& -3.93(\textbf{3.90E-02}) & -5.26(\textbf{3.90E-03}) \\
\hline
2000 & 1.66$\pm$0.23 & 1.57$\pm$0.20 & 2.08$\pm$0.30 & 0.96(1.00E+00) & -3.56(8.58E-02) & -4.54(\textbf{1.17E-02}) \\
\hline
1500 & 2.16$\pm$0.25 & 1.90$\pm$0.21 & 2.33$\pm$0.22 & 2.52(8.35E-01) & -1.66(1.00E+00)	& -4.55(\textbf{7.80E-03}) \\
\hline
1000 & 3.29$\pm$0.25 & 2.70$\pm$0.41 & 3.09$\pm$0.30 & 3.85(\textbf{4.68E-02}) & 1.59(1.00E+00) &	-2.44(9.95E-01)\\
\hline
750	&  4.32$\pm$0.37 & 3.47$\pm$0.43 & 3.92$\pm$0.53 & 4.77(\textbf{7.80E-03}) & 1.94(1.00E+00) &	-2.11(1.00E+00) \\
\hline
500	& 6.03$\pm$0.68 & 4.78$\pm$0.67 & 5.15$\pm$0.44 & 4.14(\textbf{2.34E-02}) & 3.42(1.17E-01) & -1.49(	1.00E+00) \\
\hline
250	& 11.41$\pm$1.34 & 8.92$\pm$1.14 & 8.68$\pm$0.76 & 4.46(\textbf{1.17E-02}) & 5.61(\textbf{3.90E-03}) & 0.56(1.00E+00) \\
\hline
100	& 67.51$\pm$18.79 &	22.58$\pm$1.69 & 21.20$\pm$1.40 & 7.53(\textbf{3.90E-03}) &	7.77(\textbf{3.90E-03}) &	1.98(1.00E+00) \\
\hline
50	& 64.34$\pm$7.46 & 60.42$\pm$6.83 & 57.70$\pm$4.71 & 1.23(1.00E+00) & 2.38(1.00E+00) & 1.04(1.00E+00) \\
\hline
25	& 62.98$\pm$5.59 & 58.46$\pm$5.74 & 62.70$\pm$5.84 & 1.78(1.00E+00) & 0.11(1.00E+00) & -1.64(1.00E+00)\\
\hline
10	& 100.52$\pm$10.29 & 83.90$\pm$4.48 & 84.13$\pm$4.49 & 4.68(\textbf{7.80E-03}) &	4.62(\textbf{7.80E-03}) &	 -0.11(1.00E+00) \\
\hline\hline
\end{tabular}%
}
\label{sse_ttest}
\end{table}

\subsection{Using Cross-Validation for Parameter Settings}
 \label{ct_results}
\begin{figure}
\centering
 \includegraphics[height=4.0in]{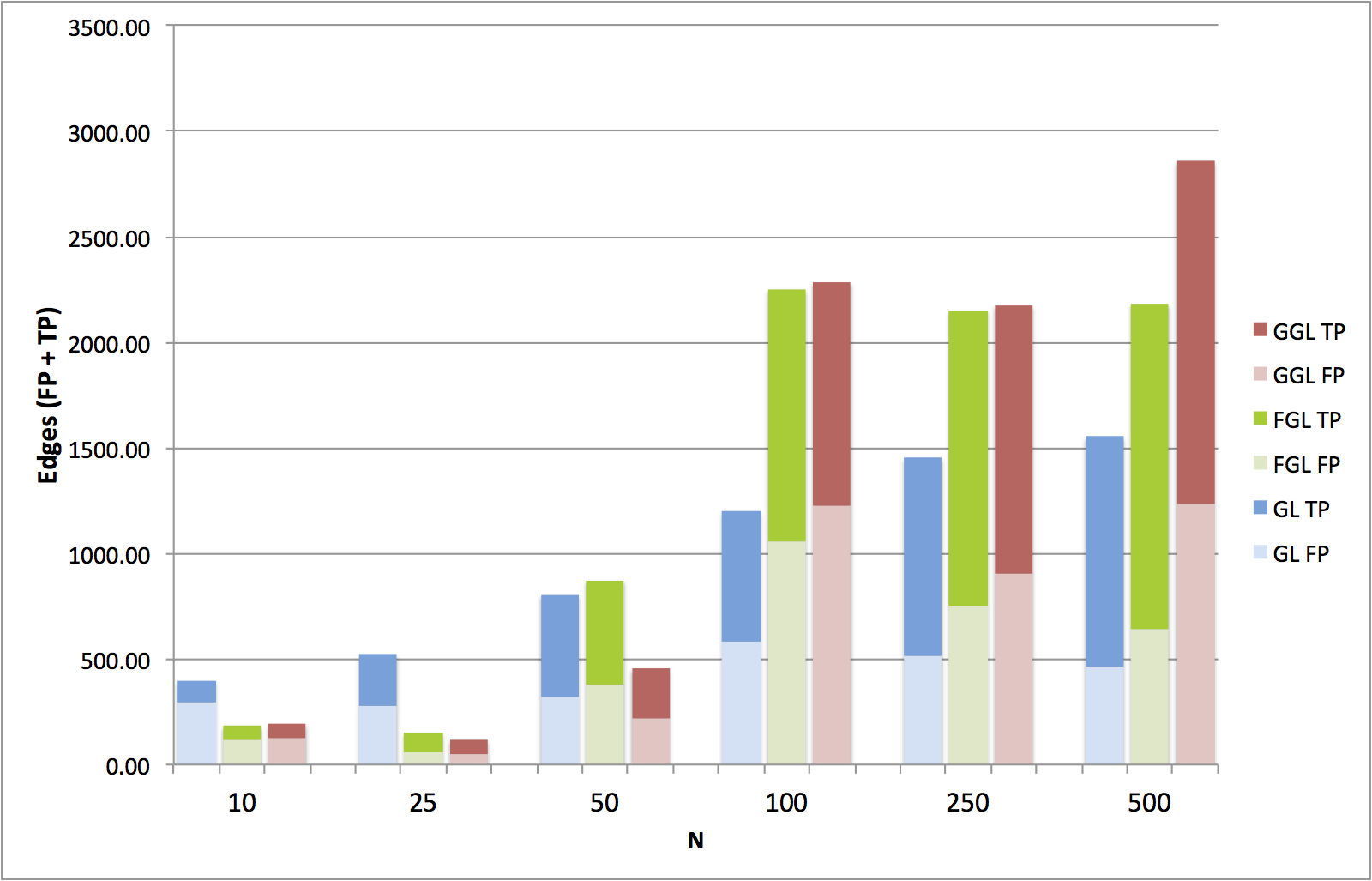}
 \caption[TP/FP edges by sample size and model for cross-validation experiment] {TP/FP edges by sample size and model for cross-validation experiment. Lower, light colored segments show the average number of FP edges whereas, upper, dark colored segments show the average number of TP edges. }
 \label{tpfp_cv}
 \end{figure}

In section \ref{gt_results} we used the precision matrices from the gold standard data set directly in determining fixed group regularization penalties (i.e., $\lambda_2$) for the joint models.  In practice, one does not typically have access to a gold standard.  In this experiment we consider how well these models perform in a practical setting, determining the regularization coefficient settings using a typical 10-fold cross validation design.   For this experiment we require 10 subsets of data, stratified by group, for regularization coefficient selection and testing with the gold standard.  We first separate the gold standard data set into two independent sets, stratified by group ($N^{(g1)}= 5,953$, $N^{(g2)}=3,775$ in each set).  From each set and group, we draw 10 subsets of images, with replacement, of equal sizes $N \in \{10,25,50,100,250,500\}$.  This procedure results in 10 subsets of data for regularization coefficient selection orthogonal to the 10 subsets for precision matrix evaluation.  Using the clusters learned from the gold standard dataset, we compute the average values for each cluster and each subject in the corresponding regularization coefficient selection subset and compute the corresponding covariance matrices between each of the clusters.  We now have 20 covariance matrices, one for each of the 10 regularization coefficient selection subsets, for each cohort, using the clusters learned from the gold standard data.  These covariance matrices will be used to determine settings for the regularization coefficients.  To set the regularization coefficients, we perform grid searches over the regularization parameter spaces (i.e., $\lambda_1$, $\lambda_2$ for FGL and GGL, and $\lambda$ for GL) for each inverse covariance model, selecting the regularization coefficient settings that minimizes the BIC score from each regularization coefficient selection subset, where the number of parameters are the number of non-zero entries in the precision matrices.  After performing this procedure, we have settings for the regularization coefficients for each of the three inverse covariance models, for each of the 10 subsets of data.  Finally, we use the gold standard clusters and the regularization coefficients to compute the precision matrices for each of the three inverse covariance models across the 10 precision matrix evaluation subsets and both cohorts.   
\\
\\
The results from this experiment are shown in Figure \ref{tpfp_cv} and Tables \ref{tpfp_practical} and \ref{sse_practical}.  Figure \ref{tpfp_cv} and Table \ref{tpfp_practical} show the average number of TP and FP connections, by model, across the 10-folds compared to the gold standard data set.  The lower segment of the stacked bars (i.e., lighter color) shows the number of FP connections whereas the upper segment (i.e., darker color) shows the number of TP connections.  For $N \in \{100,250,500\}$ the FGL and GGL models find more TP connections than GL.  For $N\in\{10,25,50\}$ GL finds more TP connections than the joint models yet all of the models are performing poorly.  The SSE between the average precision matrices across the 10 test subsets and the gold standard are shown in Table \ref{sse_practical}.   The results are similar to those of the TP/FP edges for $N\in\{100,250,500\}$, where FGL and GGL models achieve statistically lower SSEs than the GL model.  In contrast, at the lower $N$'s, there is no significant difference between the models. Interestingly, the average SSE measurements for the GL model in the larger groups (i.e., $N\in{100,250,500}$) are much higher than in the experiment described in section \ref{gt_results}; whereas, in the FGL and GGL models the results are more consistent.  In looking at the regularization parameter settings across the splits of the data for this experiment, we find the GL model is selecting higher regularization parameters (range $0.2-0.7$ depending on split and number of samples) as compared to the settings yielding the minimum average SSE when using the gold standard to set them directly (range $0.1-0.3$).  Conversely, in the FGL and GGL models, we find the $\lambda_1$ regularization parameter settings to be similar to those found yielding the minimum average SSE when using the gold standard; whereas, the $\lambda_2$, group regularization setting is generally a bit higher.  Because the group models use all the data, across both cohorts, in setting the regularization parameters, the settings are more stable across splits and result in better estimates of the connection strengths in sample sizes $N\in\{100,250,500\}$ as demonstrated in Table \ref{sse_practical}, yet in the smallest sample sizes this is not the case.          

\begin{table}[ht]
\caption{TP/FP Edges by Model and Sample Size Using Cross-Validation}
\centering
\begin{tabular}{c c c c c c c }
\hline\hline
  & GL &  & FGL  &  &   GGL  &  \\
\hline
 N & $\overline{TP}\pm$std & $\overline{FP}\pm$std &  $\overline{TP}\pm$std &  $\overline{FP}\pm$std &  $\overline{TP}\pm$std & $\overline{FP}\pm$std \\
\hline\hline
500 & 1090.6$\pm$193.1 & 464.1$\pm$153.6	& 1540.1$\pm$48.1 & 641.0$\pm$127.2 & 1625.2$\pm$24.0 & 1236.4$\pm$170.6 \\
\hline
250	& 937.9 $\pm$159.4 & 517.6$\pm$146.5 & 1400.2$\pm$35.3 & 752.8$\pm$42.9 & 1274.4$\pm$35.9 & 903.4$\pm$45.1 \\
\hline
100	& 613.2$\pm$215.5  & 585.8$\pm$257.2  & 1195.8$\pm$30.6 & 1058.2$\pm$62.2 &	1062.2$\pm$43.6 & 1223.8$\pm$141.7  \\
\hline
50  &  483.5$\pm$346.0 & 322.1$\pm$288.2	& 497.2$\pm$409.0 & 377.6$\pm$372.2 & 242.3$\pm$285.2 & 215.3$\pm$313.6 \\
\hline
25  & 239.5$\pm$193.4 & 281.6$\pm$267.9 &	91.0$\pm$111.0 & 61.8$\pm$114.2 &	69.0$\pm$107.9 & 50.9$\pm$98.1 \\
\hline
10   & 106.8$\pm$207.7 & 292.4$\pm$642.1 &	66.7$\pm$148.1 & 117.6$\pm$316.1 &	67.9$\pm$146.1 & 123.0$\pm$317.0 \\
\hline\hline
\end{tabular}
\label{tpfp_practical}
\end{table}
\begin{table}[ht]
\caption[Sum of squared errors with gold standard by model and sample size using cross-validation]{Sum of Squared Errors (SSE) with Gold Standard by Model and Sample Size Using Cross-Validation }
\centering
\resizebox{\columnwidth}{!} {%
\begin{tabular}{c c c c c c c }
\hline\hline
 & GL &  FGL  & GGL  & GL-FGL  & GL-GGL & FGL-GGL \\
\hline
N & $\overline{SSE}\pm std$ & $\overline{SSE}\pm std$ & $\overline{SSE}\pm std$ & T(P Bonferroni) & T(P Bonferroni) & T(P Bonferroni) \\
\hline\hline
500 &	156.66$\pm$70.11 &	4.88$\pm$1.04 &	8.54$\pm$3.32 &	6.85(\textbf{3.77E-05}) &	6.67(\textbf{5.26E-05}) &	-3.33(6.76E-02) \\
\hline
250 & 190.47$\pm$41.17	& 7.16$\pm$0.96 & 11.46$\pm$1.56 & 14.08	(\textbf{6.69E-10}) &	13.74(\textbf{9.98E-10}) & -7.42(\textbf{1.26E-05}) \\
\hline
100 & 207.17$\pm$46.93 &	14.28$\pm$1.64 & 31.82$\pm$16.17 &	13.00(\textbf{2.51E-09}) &	11.20(\textbf{2.85E-08}) &	-3.41(5.59E-02) \\
\hline
50	& 	266.49$\pm$75.68 & 231.15$\pm$128.5 & 332.05$\pm$91.61 &	0.75(1.00E+00) & 	-1.74(1.00E+00) &	-2.02(1.00E+00) \\
\hline
25	& 283.22$\pm$68.10 & 366.5$\pm$55.47 &	372.8$\pm$49.69 &	-3.00(1.39E-01) &	-3.36(6.27E-02) &	-0.27(1.00E+00) \\
\hline
10	& 530.03$\pm$167.41 &	342.01$\pm$41.8 &	350.91$\pm$45.33 &	3.45(5.19E-02) &	3.27(7.73E-02) &	-0.46(1.00E+00) \\
\hline\hline
\end{tabular}%
}
\label{sse_practical}
\end{table}

\section{Discussion}
\label{discussion}
In section \ref{results} we described two experiments to evaluate the inverse covariance models described in section \ref{methods}.  In section \ref{gt_results} we used the gold standard precision matrices to set the group regularization parameters in the joint models that yielded the smallest difference from the gold standard.  The results suggest that the joint models perform better in detecting true-positive functional connections and more accurately learn the connection strengths across most of the sample sizes.  At $N \in \{1500, 2000, 2500\}$ the joint models find $\sim 2-3\%$ more true positive edges than the GL algorithm,  at $N \in \{250,500,750,1000\}$ the joint models find $\sim 3-7\%$ more true positive edges than GL, and at $N \leq 100$ the joint models find $\sim 31-46\%$ more true positive edges than the GL algorithm.  The largest gain for the joint models (i.e., FGL and GGL) over the independent model (i.e., GL) in our comparisons were at $N=100$.   In section \ref{ct_results} we used a cross-validation design and orthogonal splits of the gold standard data to evaluate the performance of these models in a more practical setting, where a gold standard is not available.  The results of this experiment similarly show the joint models performing better than the GL model for data set sizes of 100 or more, with the largest difference at $N=100$.   At $N \in \{100,250,500\}$ the joint models find $\sim 26-59\%$ more true positive edges than the GL algorithm.   At $N=50$ the FGL and GL algorithms detect only $\sim 23\%$ of the total true positive edges; whereas, the GGL algorithm is much lower at $\sim 11\%$ of the true positive edges.  Even worse, at $N\leq 25$ the true positive rates for all models are $\sim 3-11\%$ which is far too low for practical use.  Further, we observe that the GL model SSE measures are much higher across the range of data set sizes when using cross validation to set the regularization weight than when using the gold standard directly (section \ref{gt_results}), caused by over regularization of the GL model in the cross-validation experiment.  The results at sample sizes less than 50 are generally quite poor and none of the models perform well using cross-validation and BIC scores to select settings for the regularization parameters.  Furthermore, this result is consistent when using cross-validation and the Akaike Information Criterion (AIC) score, a score that penalizes less for the number of model parameters than BIC, to select the regularization parameters.  Visually, we see the GL model with more true-positive edges at $N\in\{10,25\}$ than the joint models; although, all models appear to be over regularizing.  From the results of experiment 1 (see section \ref{gt_results}), using the gold standard to set the regularization parameters, we know the joint models can perform better at the smallest sample sizes, recovering $\sim 21-22\%$ of the true positive edges at $N=10$ versus $\sim 11\%$ for the GL model,  yet it is evident that care should be taken when setting the regularization parameters in the cross-validated design with such small samples.  It may be prudent to use a larger hold-out sample to set the regularization parameters prior to applying them to very small datasets.   Alternatively, one could compute the inverse covariance models across a range of regularization parameters and interpret the results based on edge selection frequencies or by controlling for false discovery rates as in \citet{liu2013gaussian}.     
\\
\\
Our results generally agree with \citet{Danaher:2013fe} who showed an improvement in the accuracy of the joint models over graphical lasso in groups with shared patterns of edges in simulated data.  In our experiments, using molecular imaging data, both joint models perform well with the FGL model moderately outperforming the GGL model in some experiments.  In practice, one should take into consideration the expected similarity of the groups being compared and the amount of available data, to determine whether the added complexity of selecting two regularization coefficients in the FGL and GGL models is warranted and to choose among the joint models.

\section{Conclusions}
\label{conclusions}
In this manuscript we have compared models for group-based functional connectivity using static molecular imaging data and given a quantitative evaluation of the models in recovering a gold standard connectivity profile, as a function of sample size.  In the largest samples, all models performed well; although, the joint models generally perform better.  In smaller samples, the joint models are more stable and achieve better results when using a large dataset to determine settings for the regularization coefficients.  Caution should be used when applying these models to very small data sets.  Our experiments suggest the true positive rates will be low and the the connection strengths will be inaccurate when using cross-validation and BIC scores to set the regularization parameters.  Given these caveats, our results show there is value in using sparse inverse covariance estimation for measuring functional connectivity in group-based molecular imaging.  It would be interesting to extend these experiments to within-subject experimental designs.  In recent work, \citet{Qiu:2013tk} has developed a joint penalty that appears to perform better than GGL in settings where there are dependencies between networks. Using experiments similar to those presented in this manuscript, an evaluation by sample size of within-subject designs would provide complimentary information on using sparse inverse covariance estimation for functional connectivity modeling in molecular imaging.

\section{Acknowledgements}
For this work we wish to thank Dr. Daniel Amen, M.D. and the Amen Clinics Inc. for supplying the gold standard dataset.   This work was supported in part by the National Center for Research Resources and the National Center for Advancing Translational Sciences, National Institutes of Health, through Grant UL1 TR000153. The content is solely the responsibility of the authors and does not necessarily represent the official views of the NIH.

%Here are two sample references: \cite{Feynman1963118,Dirac1953888}.
\pagebreak
\section{References}

\bibliography{mybibfile}

\pagebreak
\appendix
\section{Functional Clusters}
\label{clustering}
Here we describe a finite mixture model for finding functional clusters within anatomically-defined regions of interest that will subsequently be used as nodes in our network for functional connectivity modeling.  The model has been designed with respect to static SPECT and PET modalities, consisting of independent subjects across multiple cohorts, with no available anatomical scans.  Our motivation is to find a set of regions that are consistent with the observed functional data across all cohorts, while also allowing the nodes to be interpreted with respect to an anatomical reference.  From neuroanatomy, we know the brain is fundamentally composed of spatially varying clusters of cells which segregate during development into specialized units. These specialized units consist of compact clusters of neurons, forming nuclei, which contribute to the overall function of the region.  Neuroanatomic studies of brain cytoarchitectonics provides evidence for intra-regional specializations of neuronal clusters for many brain structures \cite{Anonymous:NAEtogX_,Anonymous:cnSnlYrr,Mai:1997vf}.  Using a simple average functional signal from an anatomically defined brain region over simplifies this complexity and risks missing information related to inter-regional variations in function.    
\\
\\
To learn the location and number of functional nuclei in areas constrained by an anatomical atlas, we observe that if an anatomical region has many nuclei, we would expect to find multi-modal signals within the functional imaging data.  Further, because functional imaging data is always smoothed with a Gaussian smoothing kernel during image formation, we would expect any signal from the nuclei, as measured by functional imaging, to be smoothly varying in space with a functional peak spatially close to the center-of-mass of the nuclei and an area roughly consistent with the extent of the cells composing the nuclei.  Given these observations, we propose to model the functional clusters using finite Gaussian mixture models \cite{McLachlan:1988th}.  Based on data from neuroanatomic studies, there may be multiple nuclei within many anatomically-defined regions (e.g. \citet{TONCRAY:1946tv} discuss twenty-six nuclei of the human thalamus) but we do not know how many of these nuclei result in functional signals detectable in the acquired data.  We can use this information to limit the number of clusters we search for and use model selection methods to identify how many clusters are needed to reasonably explain our observed functional signals.    
\\
\\
We begin by thinking about the imaging data as being composed of a set of voxels, each being identified by its three-dimensional coordinate  $x_{i}=(p_1,p_2,p_3)$ within the image.  For a given voxel coordinate, $x_i$, the acquired functional imaging data, is quantified to counts per unit time and provides a non-negative, real-valued proxy  $I(x_i) \in\mathbb{R}_{\geq0}$ corresponding to the number of detected events by the imaging system at that voxel coordinate. We make the simplifying assumption that each detected event is independent from all other detections and identically distributed.  This assumption implies that the functional imaging signal $I(x_i)$ at each voxel (i.e., the number of detected events) can be modeled as $I(x_i)$ independent observations at voxel $x_i$.   For each region $r$ defined in an anatomical atlas, the set of $N^{r}$ voxel coordinates contained within the boundaries of the atlas region $r = \{x_1, \dotsc , x_{N^(r)}\}$ specify the voxels assigned to the anatomical region.  Each voxel is a member of only one anatomical region.  The log-likelihood of all the detected events within a region is then modeled by a K-component Gaussian mixture model:
\begin{equation}
 \label{p_Ix_mix2}
 \ln p(X^{r} | \Theta^{\left(r\right)}) = \sum_{i=1}^{N^{(r)}} I(x_i) \ln \left\{ \sum_{k=1}^{K} w_{k}^{\left(r\right)}\mathcal{N}\left(x_i | \mu_{k}^{\left(r\right)}, \Sigma_{k}^{\left(r\right)}\right)\right\}
 \end{equation}
where $ \Theta^{\left(r\right)}= \{ w_{1}^{\left(r\right)}, \dotsc ,w_{k}^{\left(r\right)}, \mu_{1}^{\left(r\right)}, \Sigma_{1}^{\left(r\right)}, \dotsc, \mu_{k}^{\left(r\right)}, \Sigma_{k}^{\left(r\right)} \}$ are the model parameters, $X^{r}$ is the set of all voxel coordinates contained in region $r$, and $I(x_i)$ is the number of detected events at the voxel coordinate $x_i$ for all $N^{(r)}$ coordinates.  Parameters $\mu_{k}^{(r)}$ and $\Sigma_{k}^{\left(r\right)}$ represent the mean spatial location and covariance matrix of the $k^{th}$ cluster in region $r$.  The mixing weights $w_{k}^{(r)}$ represent the probability that $x_i$ was generated by component $k$ and are subject to the constraints  $\sum_{k=1}^{K} w_{k}^{\left(r\right)}= 1$ and $0 \leq  w_{k}^{\left(r\right)} \leq 1$ to be valid probabilities.  The unknown parameters $ \Theta^{\left(r\right)}$ can be learned using the iterative expectation maximization (EM) algorithm and maximized in closed form \cite{Dempster:1977fi, Krishnan:1997uh}.  Although the algorithm is guaranteed to converge, it is not guaranteed to converge to a global maximum of the likelihood function and is initialization dependent.  To initialize the mixture model we use the kmeans++ algorithm of \citet{Arthur:2007tv}, an extension to the K-means algorithm.  K-means is a clustering technique which finds a partitioning of the data into $K$ clusters that minimizes the sum of the squared distances between each data point and its closest cluster center and is routinely used to initialize Gaussian mixture models.  The kmeans++ algorithm is a simple extension that uses a random seeding technique which has been shown to improve speed and accuracy with respect to the recovering the true cluster centers.  To initialize our Gaussian mixture models using this technique, we choose an initial cluster center from the data for a particular anatomical region with the highest number of detected events.  We then choose the next cluster center at random from the remaining data points with probability proportional to the distance of each data point to the closest previously selected cluster center, weighted by the number of detected events.   This step gives higher probability to data points farther away from the previously selected cluster centers, weighted by the number of detections at those points.   The procedure continues until all initial $K$ cluster centers have been selected.  We then run the k-means algorithm to convergence and use the assignments of the voxels to the $K$ clusters to initialize our Gaussian mixture model means and covariances.
\\
\\
The final step in the clustering model is to select an appropriate setting for the number of clusters $K$ for each region.  We cannot simply choose the model that explains the data best because more complex models (i.e., larger $K$) will always explain the data better than a simplier model with fewer clusters.  Among the various approaches to model selection, penalized likelihood methods have been shown to be competitive, simple to implement in practice, and to work well when applied to mixture models \cite{Smyth:2000hk}.  Here we choose the Bayesian information criterion (BIC) developed by \citet{Schwarz:1978kf} for model selection, which has been shown to be a consistent estimator for independent and identically distributed observations in linear exponential family models such as the Gaussian mixture model  \cite{Haughton:1988is}.  By choosing the model with the minimum BIC score, one is attempting to select the candidate model with the highest Bayesian posterior probability.   The BIC score penalizes the model likelihood and tends to favor simple models over more complex ones.  The BIC score for our mixture model within a brain region is given by:  
\begin{equation}
 \label{p_Ix_mix}
BIC\left(\Theta^{\left(r\right)}\right) =  -2 \ln p\left(X | \Theta^{\left(r\right)}\right) + \rho^{\left(r\right)} \ln \left(\sum_{i=1}^{N^{(r)}} I(x_i)\right) 
 \end{equation}
where $\rho^{\left(r\right)} $ is the number of parameters in the model for region $r$ and is used to penalize the likelihood relative to model complexity.  In our model we are learning $K$ three-dimensional Gaussian distributions for each region.  For each Gaussian distribution of dimension $D=3$, we need $\rfrac{D}{2}(D+1)$ parameters to specify the symmetric covariance matrix and $D$ parameters to specify the mean.  There are $K$ mixing weights for the K-component Gaussian mixture model, resulting in $K*\rfrac{D}{2}(D+3) + K$ parameters.  After learning the mixture models for each setting of $K$, we compute the BIC score, penalizing the likelihood scores by the number of parameters estimated.  We select the best model as the one with the lowest BIC score, which implies the model with the highest relative data likelihood after penalizing for model complexity.  Once the best models have been selected for each brain region, we compute values for each cluster by averaging over the number of detections at each voxel, weighted by their cluster membership probabilities.  We use these cluster averages, computed for each subject, in the sparse inverse covariance models.

\section{Inverse covariance model results without clustering}
The results showing the average area under the TP/FP edge curves (AUC) across the subsets, by sample sizes, using the gold standard for parameter settings as described in section \ref{gt_results} but without the clustering model are shown in Table \ref{auc_ttest_nocluster}.  The functional data was averaged in each region of interest instead of using the results from the Gaussian mixture model described in \ref{clustering}.  The AUCs are statistically lower in the GL model (i.e. negative t-scores) than both the FGL and GGL models at data set sizes from 50-2000, with the moderate sample sizes showing the largest decreases.  These results are consistent with those when using the clustering model, yet the differences between inverse covariance models are lower in magnitude and the AUCs are slightly higher, likely due to the smaller number of variables relative to sample size.
\\
\\
The results showing the average minimum SSEs across the subsets, by sample sizes, using the gold standard for parameter settings as described in section \ref{gt_results} but without the clustering model are shown in Table \ref{sse_ttest_nocluster}.  Again, the results are consistent with those presented using the clustering model, but the SSEs are less accurate over the range of sample sizes likely related to averaging functional signals with noise in the larger regions of interest.   

\begin{table}[ht]
\caption[TP/FP edge by model and sample size using ROI averaging]{TP/FP Edge Area Under Curve (AUC) T-test comparison by model and sample size using gold standard for parameter settings and simple averaging of functional signal in AAL regions of interest}
\centering
\resizebox{\columnwidth}{!} {%
\begin{tabular}{c c c c c c c }
\hline\hline
 & GL  &  FGL   & GGL  &GL-FGL  & GL-GGL & FGL-GGL \\
\hline
N & $\overline{AUC}\pm std$ & $\overline{AUC}\pm std$ & $\overline{AUC}\pm std$ & T(P Bonferroni) & T(P Bonferroni) & T(P Bonferroni) \\
\hline\hline
2500 &	0.840$\pm$0.016 &	0.851$\pm$0.016 &	0.856$\pm$0.011 &	-1.538(1.00E+00) &	-2.603(7.01E-01) &	0.814(1.00E+00) \\
\hline
2000 &	0.823$\pm$0.009 &	0.842$\pm$0.011 &	0.842$\pm$0.010 &	-4.157(\textbf{2.31E-02}) &	-4.388(\textbf{1.38E-02}) &	-0.043(1.00E+00) \\
\hline
1500	 & 0.806$\pm$0.011 &	0.828$\pm$0.010 &	0.824$\pm$0.007 &	-4.517(\textbf{1.04E-02}) &	-4.421(\textbf{1.29E-02}) &	-0.773(1.00E+00) \\
\hline
1000	 & 0.772$\pm$0.009 &	0.801$\pm$0.011 &	0.796$\pm$0.009 & -6.398(\textbf{1.96E-04}) &	-5.937(\textbf{5.00E-04}) &	-1.184(1.00E+00) \\
\hline
750 & 0.751$\pm$0.009 &	0.782$\pm$0.009 &	0.775$\pm$0.007 &	-7.511(\textbf{2.32E-05}) &	-6.623(\textbf{1.26E-04}) &	-1.933(\textbf{1.00E+00}) \\
\hline
500	& 0.720$\pm$0.011 &	0.753$\pm$0.012 &	0.742$\pm$0.009 &	-6.426(\textbf{1.86E-04}) &	-5.053(\textbf{3.23E-03}) &	-2.192(1.00E+00) \\
\hline
250 & 0.656$\pm$0.011 &	0.682$\pm$0.014 &	0.675$\pm$0.013 &	-4.600(\textbf{8.67E-03}) &	-3.505(\textbf{9.85E-02}) &	-1.195(1.00E+00) \\
\hline
100 & 0.617$\pm$0.011 &	0.638$\pm$0.011 & 	0.631$\pm$0.009 &	-4.352(\textbf{1.50E-02}) &	-3.177(2.04E-01) &	-1.531(1.00E+00) \\
\hline
50 & 0.576$\pm$0.014 & 	0.604$\pm$0.013 &	0.597$\pm$0.011 &	-4.502(\textbf{1.07E-02}) & 	-3.708(\textbf{6.27E-02}) &	-1.189(1.00E+00) \\
\hline
25 & 0.560$\pm$0.014 &	0.576$\pm$0.011 &	0.574$\pm$0.008 &	-2.869(3.98E-01) &	-2.622(6.75E-01) &	-0.591(1.00E+00) \\
\hline
10 & 0.547$\pm$0.013 &	0.559$\pm$0.015 &	0.552$\pm$0.010 &	-1.878(1.00E+00) &	-0.945(1.00E+00) &	-1.199(1.00E+00) \\
\hline\hline
\end{tabular}%
}
\label{auc_ttest_nocluster}
\end{table}

\begin{table}[ht]
\caption[Sum of squared errors by model and sample size using ROI averaging]{Minimum average sum of squared errors (SSE) by model and sample size using gold standard for parameter settings and simple averaging of functional signal in AAL regions of interest}
\centering
\resizebox{\columnwidth}{!} {%
\begin{tabular}{c c c c c c c}
\hline\hline
& GL  &  FGL   & GGL &GL-FGL  & GL-GGL & FGL-GGL \\
\hline
N & $\overline{SSE}\pm std$ & $\overline{SSE}\pm std$ & $\overline{SSE}\pm std$ & T(P Bonferroni) & T(P Bonferroni) & T(P Bonferroni) \\
\hline\hline
2500 &	5.94$\pm$5.08	& 7.13$\pm$6.11 &	9.59$\pm$8.30 &	-0.472(1.00E+00) &	-1.184(1.00E+00) &	-0.753(1.00E+00) \\
\hline
2000	 & 7.69$\pm$8.88 &	9.07$\pm$7.79 &	14.36$\pm$12.14 &	0.397(1.00E+00) &	-0.800(1.00E+00) &	-1.159(1.00E+00) \\
\hline
1000	 & 17.39$\pm$14.67 &	11.03$\pm$9.49 &	22.79$\pm$19.20 &	1.150(1.00E+00) &	-0.707(1.00E+00) &	-1.736(1.00E+00) \\
\hline
750 & 21.90$\pm$18.63 &	13.19$\pm$11.34 &	25.88$\pm$22.59 &	1.263(1.00E+00) &	-0.430(1.00E+00) &	-1.588(1.00E+00) \\
\hline
500 & 33.68$\pm$28.56 &	19.42$\pm$17.05 &	23.84$\pm$20.19 &	1.356(1.00E+00) &	0.890(1.00E+00) &	-0.529(1.00E+00) \\
\hline
250 & 44.23$\pm$37.35 &	42.54$\pm$37.06 &	41.11$\pm$34.53 &	0.102(1.00E+00) &	0.194(1.00E+00) &	0.089(1.00E+00) \\
\hline
100 & 78.07$\pm$67.43 &	60.83$\pm$51.49 &	69.76$\pm$60.06 &	0.643(1.00E+00) &	0.291(1.00E+00) &	-0.357(1.00E+00) \\
\hline
50 & 116.36$\pm$98.85 &	94.76$\pm$79.63 &	101.89$\pm$85.90 &	0.538(1.00E+00) &	0.350(1.00E+00) &	-0.192(1.00E+00) \\
\hline
25 & 188.52$\pm$159.39 & 	165.43$\pm$140.50 &	180.41$\pm$153.59 &	0.344(1.00E+00) &	0.116(1.00E+00) &	-0.228(1.00E+00) \\
\hline
10 & 241.64$\pm$204.92 &	224.41$\pm$189.45 &	231.57$\pm$195.92 & 	0.195(1.00E+00) &	0.112(1.00E+00) &	-0.083(1.00E+00) \\
\hline\hline
\end{tabular}%
}
\label{sse_ttest_nocluster}
\end{table}

\end{document}